\documentclass[12pt,draftclsnofoot,onecolumn]{IEEEtran}
\usepackage{amsmath, amssymb}
\usepackage{tipa}
\usepackage{textgreek}
\usepackage[caption=false,font=normalsize,labelfont=sf,textfont=sf]{subfig}
\usepackage{flushend}
\usepackage{mathtools}  
\usepackage[table,xcdraw]{xcolor}
\usepackage[sort]{cite}

\begin{document}

\pagenumbering{gobble}
\title{\huge{Statistical Inference-Based Channel Estimation for LD-Driven Visible Light O-OFDM Systems in the Presence of Relative Intensity and Input-Signal-Dependent Shot Noise}}
\author{
Shubham~Saxena,~\IEEEmembership{Graduate Student Member,~IEEE,} Karra~Sreeman~Reddy, Suraj~Srivastava,~\IEEEmembership{Member,~IEEE,} Subrahmanya~Swamy~Peruru, ~\IEEEmembership{Member,~IEEE,} Aditya~K.~Jagannatham,~\IEEEmembership{Senior Member,~IEEE,} and Lajos~Hanzo, ~\IEEEmembership{Life Fellow,~IEEE}\vspace{-10mm} 
\thanks{Shubham Saxena, Karra Sreeman Reddy, Subrahmanya Swamy Peruru, and Aditya K. Jagannatham are with the Department of Electrical Engineering, Indian Institute of Technology Kanpur, Kanpur-$208016$, India (e-mail: \{shubs20; sreemanrk23; swamyp;  adityaj\}@iitk.ac.in). Suraj Srivastava is with the Department of Electrical Engineering, Indian Institute of Technology Jodhpur, Rajasthan $342030$, India (email: surajsri@iitj.ac.in). L. Hanzo is with the School of Electronics and Computer Science, University of
Southampton, Southampton SO$17$ $1$BJ, U.K. (email: lh@ecs.soton.ac.uk).
}
}
\date{\today}
\maketitle
\vspace{-10mm}
\begin{abstract}
Laser diode (LD)-based luminaires are gaining increasing attention in automotive applications and are expected to extend to residential and commercial environments, creating opportunities for high-bandwidth visible light communication (VLC) systems. However, practical LD-based VLC links are impaired by input-signal-dependent shot noise (ISDSN), relative intensity noise (RIN), and thermal noise, which affect reliable channel estimation (CE). This work investigates their joint impact on receiver-side CE in a single-input single-output (SISO) optical orthogonal frequency division multiplexing (O-OFDM) VLC system under a statistically random channel model. A statistical inference framework is developed in which the receiver exploits observed signal variations to estimate the channel under optical impairments. Closed-form expressions are derived for least squares (LS), maximum likelihood (ML), maximum a posteriori probability (MAP), minimum mean square error (MMSE), and linear MMSE (LMMSE) estimators. In addition, the Bayesian Cramér-Rao lower bound (BCRLB) is derived to benchmark mean square error (MSE) performance. Monte Carlo simulations for direct current-biased O-OFDM (DCO-OFDM) and asymmetrically clipped O-OFDM (ACO-OFDM) validate the analysis. Results show substantial CE degradation under the joint presence of ISDSN and RIN, while the MMSE estimator consistently achieves the lowest MSE, demonstrating strong potential for robust and adaptive receiver operation in practical VLC systems.
\end{abstract}
\vspace{-4mm}
\vspace{-6mm} 
\section{Introduction}
\IEEEPARstart{R}{ising} demand for high-speed wireless data, driven by the proliferation of Internet of Things (IoT) devices, pervasive smartphone use, and the continued densification of mobile access networks, has intensified radio-frequency (RF) spectrum scarcity and raised concerns of an imminent spectrum crunch. To alleviate RF congestion, future wireless architectures are expected to incorporate optical wireless communication (OWC) as a complementary technology, with visible light communication (VLC) constituting a particularly promising candidate for short-range high-rate links \cite{bai2022low,ahmad2021energy,pham2024design}. In this broader context, VLC has attracted growing interest as a complementary wireless access solution for indoor networking, RF-sensitive environments, and short-range infrastructure-assisted communication scenarios. VLC operates in the visible optical band and may rely on photodiodes (PDs), laser diodes (LDs), and light-emitting diodes (LEDs), enabling simultaneous illumination and data transfer. The visible band is roughly $10~000$ times wider than the RF band, it is largely unregulated, produces negligible electromagnetic radiation, and is resilient to electromagnetic interference. These attributes make VLC attractive for dense indoor deployments and for environments where RF emissions are restricted. VLC also enhances physical-layer security since visible light cannot penetrate opaque barriers, thereby limiting eavesdropping. Moreover, leveraging the existing lighting infrastructure enables energy- and cost-efficient communication consistent with green communication objectives \cite{saxena2025multiple,li2016user,cang2022optimal}.

In indoor VLC deployments, commercially available LEDs have achieved multi-Gigabit-per-second rates, but their limited modulation bandwidth often requires high-order modulation and sophisticated equalization. Micro-LEDs offer increased bandwidth; however, their low optical output power limits coverage, and high carrier injection densities introduce efficiency penalties \cite{xie2025serially,yaseen2024signal,yaseen2023channel}. LD-based luminaires therefore, constitute a promising alternative for high-speed indoor VLC and high-capacity short-range OWC links. Compared with LEDs, LDs provide higher modulation bandwidth and improved spectral efficiency, but they are affected by both signal-independent and signal-dependent noise impairments \cite{xie2025serially,yaseen2024signal,yaseen2023channel}. Thermal noise is the dominant signal-independent impairment and is commonly modeled as additive Gaussian noise \cite{gao2016modulation}. Signal-dependent noise includes input-signal-dependent shot noise (ISDSN) and relative intensity noise (RIN) \cite{xie2025serially,yaseen2024signal,yaseen2023channel}. ISDSN arises from photon arrival randomness and can significantly degrade channel estimation (CE) accuracy, thereby increasing bit-error-rate (BER) \cite{xie2025serially,yaseen2024signal,yaseen2023channel}. RIN results from LD output power fluctuations around its mean, is well approximated by a Gaussian model, and reduces the signal-to-noise ratio (SNR) as its variance scales with signal power \cite{yaseen2024signal}. Therefore, reliable CE in LD-based OWC/VLC links requires explicit treatment of both signal-independent and signal-dependent noise processes. Such observation-driven characterization is also central to intelligent receivers, which perceive channel and noise conditions and adapt inference accordingly.

In VLC, intensity modulation with direct detection (IM/DD) is typically adopted, requiring real-valued and non-negative transmitted waveforms \cite{ghassemlooy2019optical}. Common single-carrier schemes such as pulse-width modulation, on-off keying, and $M$-ary pulse-amplitude modulation often require multitap equalization at high data rates, increasing receiver complexity \cite{ghassemlooy2019optical}. Moreover, the VLC channel comprises both line-of-sight (LoS) and non-LoS (NLoS) paths. Prior studies \cite{chen2016adaptive,zhao2014compressed,schulze2016frequency,schulze2024dispersive} have modeled VLC multipath behavior, while \cite{zhou2014impact} has shown that higher-order reflections significantly affect high-rate performance, indicating that single-reflection models are inadequate. Accordingly, this treatise adopts a multipath VLC channel model incorporating both LoS and NLoS components. Multipath propagation imposes temporal dispersion, as characterized by the power delay profile in \cite{lee2011indoor}, and the resulting delay spread causes intersymbol interference (ISI) in indoor VLC systems \cite{9223722,barry1993simulation}.

Consequently, optical orthogonal frequency division multiplexing (O-OFDM) has emerged as a promising modulation technique for VLC systems due to its inherent capability to mitigate ISI while maintaining high spectral efficiency \cite{van2021deep,h1,popoola2014pilot}. Direct current-biased O-OFDM (DCO-OFDM) and asymmetrically clipped O-OFDM (ACO-OFDM) are the most widely used O-OFDM variants in VLC \cite{ghassemlooy2019optical}. DCO-OFDM adds a sufficiently high positive direct current (DC) bias to generate a positive time-domain signal, whereas ACO-OFDM achieves unipolarity by clipping negative samples and transmitting only positive ones. As a result, ACO-OFDM is more energy-efficient, while DCO-OFDM offers higher spectral efficiency \cite{ghassemlooy2019optical}. However, both suffer performance degradation due to ISDSN, RIN, and multipath-induced dispersion. Accurate CE must therefore explicitly account for these impairments, and O-OFDM receivers require robust equalization supported by reliable CE. This motivates intelligent physical-layer processing in which the receiver perceives channel observations, learns the underlying impairments, and adaptively refines estimation for improved communication reliability. Accordingly, we develop a set of five estimators for LD-driven DCO-OFDM and ACO-OFDM VLC systems under ISDSN, RIN, and thermal noise, assuming a statistically random channel model. Related studies are critically appraised in the following subsection.

\vspace{-3mm}
\subsection{State-of-the-art}

Despite its advantages, VLC cannot fully realize its potential in practical short-range OWC systems without accurate CE \cite{zhang2018multi}. Owing to the combined LoS and NLoS multipath components arising from reflections, robust VLC system design necessitates efficient yet precise CE. Such CE problems are particularly relevant in short-range wireless access scenarios supported by optical front-ends, where VLC serves as a complementary OWC solution and the receiver must reliably infer the underlying channel from observed signals. In \cite{zhao2013channel}, least square (LS) and linear minimum mean square error (LMMSE) techniques were applied for CE in O-OFDM-VLC systems, assuming a fixed channel impulse response (CIR) length $L_h = 6$ and signal-independent noise. Further, Mohapatra \textit{et al.} \cite{mohapatra2020performance} derived a closed-form expression for symbol error probability in a single-input-single-output (SISO) VLC system under imperfect channel state information (CSI), utilizing LS estimation. Pal \textit{et al.} \cite{pal2022channel} considered LS-based CE in VLC systems with a wide field-of-view (FoV) and random receiver orientation and placement, again under the assumption of signal-independent noise. The study in \cite{gao2016modulation} introduced power-efficient modulation schemes for both single-carrier and multi-carrier VLC systems under practical lighting constraints, while assuming a deterministic channel model and accounting only for ISDSN. A blind CE strategy under similar noise conditions was developed in \cite{gurbilek2023blind}, whereas \cite{saxena2025multiple,saxena2023sparse} explored Bayesian learning-based CE for O-OFDM-VLC systems, albeit only under signal-independent noise scenarios.

\begin{table*}[!ht]
\centering
\caption{Boldly contrasting our contributions to the literature}
\label{tab:feature-comparison}
\resizebox{0.9\linewidth}{!}{%
\begin{tabular}{|l|c|c|c|c|c|c|c|c|c|}
\hline
\textbf{Features} & \cite{cang2022optimal} & \cite{yaseen2024signal} & \cite{yaseen2023channel} & \cite{gao2016modulation} & \cite{saxena2023sparse} & \cite{mitra2018minimum} & \cite{yaseen2021visible} & \cite{cheema2021distance}  & \textbf{Proposed} \\
\hline
LoS channel model           & \checkmark     & \checkmark & \checkmark & \checkmark & \checkmark & \checkmark & \checkmark & \checkmark & \checkmark \\ \hline
ISDSN system model         & \checkmark      & \checkmark & \checkmark & \checkmark &            &            & \checkmark & \checkmark & \checkmark \\ \hline
NLoS channel model       &        &            &            & \checkmark & \checkmark &            &            &                        & \checkmark \\ \hline
DCO-OFDM                   &      &            &            & \checkmark & \checkmark &            &            &            &             \checkmark \\ \hline
ACO-OFDM               &          &            &            &            & \checkmark &            &            &            &             \checkmark \\ \hline
\textbf{LS-based O-OFDM CE}    &   &            &            &            & \checkmark &            &            &            &             \checkmark \\ \hline
\textbf{LMMSE-based O-OFDM CE}   &  &            &            &            & \checkmark &            &            &            &             \checkmark \\ \hline
\textbf{O-OFDM-based BCRLB}       &        &            &            &            & \checkmark &            &            &            &             \checkmark \\ \hline
\textbf{RIN system model}        &         & \checkmark &            &            &            &            &            &            &             \checkmark \\ \hline
\textbf{ML-based O-OFDM CE}    &    &            &            &            &            &            &            &            &             \checkmark \\ \hline
\textbf{MAP-based O-OFDM CE}   &    &            &            &            &            &            &            &            &             \checkmark \\ \hline
\textbf{MMSE-based O-OFDM CE}  &   &            &            &            &            &            &            &            &             \checkmark \\ \hline
\textbf{FD statistical channel model}  &   &            &            &            &            &            &            &            &             \checkmark \\
\hline
\end{tabular}%
}\vspace{-3mm}
\end{table*}
However, these studies do not jointly incorporate signal-dependent noise impairments and statistical channel modeling, both of which are critical for reliable high-rate OWC/VLC links and for intelligent receiver operation under uncertainty. Recent studies such as \cite{yin2016performance} considered a statistical VLC model to analyze a non-orthogonal multiple access (NOMA) system, including derivations for the coverage probability and ergodic sum rate under LoS components only. In \cite{mitra2018minimum}, a minimum error entropy-based CE approach was developed based on the probability density function (PDF) of the LoS VLC gain. In another work \cite{jani2019performance}, a cooperative dual-hop power line communication (PLC)-VLC system was proposed, where the VLC link interfaced with a PLC backhaul via decode-and-forward relaying, again using the LoS gain PDF for SNR analysis. Nevertheless, these studies do not address CE in O-OFDM VLC systems under jointly realistic channel and noise assumptions. In practical LD-based VLC links, accurate CE further requires explicit treatment of both signal-dependent and signal-independent impairments. In this regard, \cite{yaseen2021visible} examined LS- and maximum likelihood (ML)-based CE under deterministic SISO VLC channels influenced by ISDSN, establishing baseline mean-square-error (MSE) performance. Xie \textit{et al.} \cite{xie2025serially} proposed serially connected bidirectional expectation propagation (EP)-based decision feedback equalization in a multiple-input multiple-output (MIMO)-VLC system, considering only the LoS case and ISDSN. Further, \cite{yaseen2023channel} assessed LS, ML, maximum \textit{a posteriori} probability (MAP), minimum mean square error (MMSE), and LMMSE under an ISDSN-influenced statistical SISO VLC channel. The impact of ISDSN on distance estimation error bounds was studied in \cite{cheema2021distance}, while \cite{cang2022optimal} investigated the downlink of a NOMA-VLC system under ISDSN alone. The joint effects of RIN and ISDSN were studied in \cite{yaseen2024signal} within a deterministic framework using LS and ML estimators. However, to the best of the authors’ knowledge, no existing study addresses the combined influence of RIN, ISDSN, and thermal noise in an O-OFDM VLC system under a statistically random channel. This gap is particularly significant for reliable communication in practical systems, where robust receiver functionality relies on perception-driven and adaptive inference under realistic impairments. This identified gap therefore constitutes the primary motivation for the present investigation. Table \ref{tab:feature-comparison} contrasts our contributions with the existing literature. The key contributions of this work are detailed as follows.

\vspace{-4mm}
\subsection{Contributions}
\begin{enumerate}
    \item We analyze the combined impact of RIN and ISDSN, in conjunction with thermal noise, on the performance of an LD-based indoor DCO-OFDM and ACO-OFDM VLC system in the presence of a random channel. In this work, RIN is modeled as being directly proportional to the transmitted optical power and the channel gain, whereas ISDSN is considered to scale with the square root of both parameters \cite{yaseen2024signal}. This model provides a realistic representation of the impairments encountered by practical VLC receivers under dynamic operating conditions.

\item A closed-form prior PDF is derived for the composite LoS and diffuse NLoS per-subcarrier channel frequency response (CFR), with the LoS PDF obtained under random user locations and the diffuse NLoS term modeled via a deterministic room-averaged approach. This statistical characterization enables principled channel inference in practical VLC scenarios.

\item To evaluate the theoretical performance bounds of the proposed estimators, we derive the Fisher information for the considered system model, which enables the computation of the Bayesian Cram\'er-Rao lower bound (BCRLB).

\item Closed-form expressions for the MSE corresponding to five distinct estimators, namely, LS, ML, MAP, MMSE, and LMMSE, are analytically derived, and their accuracy is validated through numerical evaluation.

\item We present simulation results that highlight the substantial degradation in system performance caused by the presence of RIN and ISDSN in addition to thermal noise in the DCO-OFDM and ACO-OFDM VLC system model. Among the proposed estimators, the MMSE-based estimator consistently attains the lowest MSE, thereby highlighting its effectiveness for robust receiver-side channel inference.
\end{enumerate}
\subsection{Organization of the paper}
The rest of this paper is arranged as follows. Section II introduces the DCO-OFDM and ACO-OFDM VLC system models under the combined influence of ISDSN, RIN, and thermal noise. Section III details the multipath VLC CFR model. Section IV derives the BCRLB and corresponding Fisher information for specific noise scenarios. Section V analyzes the performance of five CFR estimation methods: LS, LMMSE, ML, MAP, and MMSE. Section VI presents simulation results evaluating the proposed CE techniques, and Section VII concludes the paper. 

\textbf{Notations:} The notations used are as follows: FD vectors are represented using bold lowercase letters. The operator $\mathbb{E} \{ \cdot \}$ represents expectation, diag$\{ \boldsymbol{a} \}$ indicates a diagonal matrix with the vector $\boldsymbol{a}$ on the principal diagonal, while the transpose of a matrix is indicated by \( [\cdot]^T \). A Gaussian random variable with mean \( \mu \) and variance \( \sigma^2 \) is given as \( \mathcal{N}(\mu, \sigma^2) \). The hypergeometric function \( {}_2F_1(a,b;c;z) \) is defined through its series expansion as \( {}_2F_1(a,b;c;z) = \sum_{i=0}^{\infty} \frac{(a)_n (b)_n}{(c)_n} \frac{z^n}{n!} \), where $(a)_n = a(a+1)...(a+n-1)$ with $(a)_0=1$. The norm of a vector \( \boldsymbol{x} \) is given by \( ||\boldsymbol{x}|| = \sqrt{\boldsymbol{x}^T \boldsymbol{x}} \), and \( f_x(x) \) denotes the PDF of the random variable \( x \). The identity matrix of size $N$ is given as $\mathbf{I}_N$. The quantity $\tilde{(\cdot)}$ denotes a variable in the time-domain (TD), $\overline{(\cdot)}$ is a vector in TD, and the rect($x$) function is defined as follows, $\text{rect}(x) = 1$, if $|x| \leq 1$, otherwise $0$.

\begin{figure*}
\centering
\includegraphics[width=0.9\linewidth,height = 60mm]{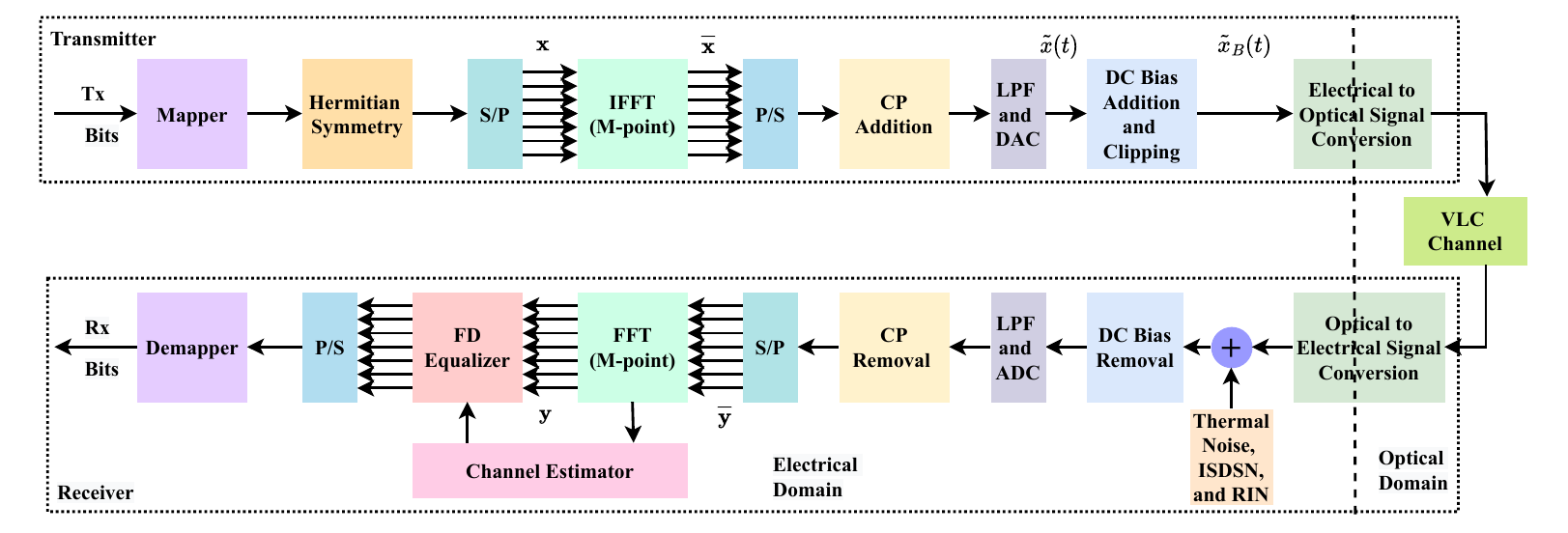}
\caption{Schematic diagram of a DCO-OFDM transmitter and receiver with ISDSN and RIN-based VLC system.}
\label{DCO1}
\end{figure*}
\section{O-OFDM ISDSN and RIN VLC System Models}
The two most often used O-OFDM systems, DCO-OFDM and ACO-OFDM, are detailed in this section. Unlike RF-based OFDM systems, DCO-OFDM and ACO-OFDM systems directly modulate the illumination of the LDs. Hence, the output signal must be real-valued and non-negative.
\subsection{DCO-OFDM system model}
The schematic representation of the DCO-OFDM system incorporating ISDSN and RIN-based transceiver components is illustrated in Fig. \ref{DCO1}.
Initially, the serial bit sequence of a single OFDM block with $M$ subcarriers is converted to complex-valued symbols via quadrature amplitude modulation (QAM). The resultant modulated signal is hosted by: $\mathbf{{x}} = [{x}_0, \hdots, {x}_{M-2}, {x}_{M-1}]^T\in \mathbb{C}^{M \times 1}$, so that the information symbols are located from ${x}_1 ~\text{to}~ {x}_{M/2 - 1}$ and follow the characteristic, ${x}_j = {x}^{\ast}_{M-j}, ~ \text{where} ~ M/2 +1 \leq j \leq M-1$. Moreover, to eliminate the DC signal, we set ${x}_0 = {x}_{M/2} = 0$. The real-valued TD signal of the vector $\mathbf{{x}}$ is determined using the $M$-point inverse fast Fourier transform (IFFT), given as $\mathbf{\overline{x}} = \mathbf{W} \mathbf{x}$, where $\mathbf{W} = \{ w_{n,k} \}^{M-1}_{n,k = 0} \in \mathbb{C}^{M \times M} $ and $w_{n,k} = \frac{1}{M}e^{j \left(\frac{2\pi n k}{M} \right)}$. Consequently, the $n^{th}$ TD symbol $\tilde{x}_n$ of the vector $\mathbf{x}$ is defined as follows \cite{saxena2023sparse}:
\begin{equation}
\tilde{x}_{n}=\frac{1}{M} \sum_{k=0}^{M-1} {x}_k  e^{\left(j \frac{2 \pi k n}{M}\right)}=\frac{2}{M} \sum_{k=1}^{M / 2-1} \operatorname{Re}\left({x}_k e^{\left(j \frac{2 \pi k n}{M}\right)}\right).
\end{equation}
The resultant signal undergoes parallel-to-serial (P/S) conversion, followed by the concatenation of a cyclic prefix (CP). With the objective of eliminating the ISI, the length of the CP ($L_{\text{CP}}$) is set higher than the VLC multipath channel's delay spread. The waveform $\tilde{x}(t)$ is then produced by the digital-to-analog converter (DAC) and subsequently applied to a low-pass filter (LPF). For intensity modulation, $\tilde{x}(t)$ must be unipolar in nature. Therefore, a DC bias ($B_{\text{DC}}$) is added to $\tilde{x}(t)$, which is given by $B_{\text{DC}}=  \text{\textscriptv} \sqrt{\mathbb{E} \{ \tilde{x}^2(t)\} }$, and the constant \textscriptv ~is chosen to satisfy $B_{\text{DC}} = 10\log(\text{\textscriptv}^2+1)$ \cite{kahn1997wireless}. The resultant unipolar signal is described as $\tilde{x}_B(t) = \tilde{x}(t) + B_{\text{DC}}$. Consequently, the remaining negative signal contribution is forced to zero, due to the introduction of $B_{\text{DC}}$. The signal obtained is subsequently converted into an optical signal, followed by transmitting it through the multipath VLC channel described by the $L_h$-length vector $\mathbf{\overline{h}} = \left[ \tilde{h}(0), \tilde{h}(1), \tilde{h}(2), \hdots,\tilde{h}(L_h-1)\right]^T \in \mathbb{R}^{L_h \times 1}_{+}$. The receiver side processing includes CP elimination, converting serial data to parallel (S/P), and the fast Fourier transform (FFT). \textcolor{black}{To obtain an analytically tractable representation, we adopt the white scalar approximation, which is widely used in O-OFDM and VLC systems and becomes accurate when the number of subcarriers is large \cite{huang2023performance,tsonev2013complete,dimitrov2012information}. Thus, the input-output model for the $k^{th}$ subcarrier will be \cite{yaseen2024signal}:
\begin{equation}\label{eq5}
y_k=x_kh_k+\sqrt {|x_kh_k|}w_{{sd,k}}+x_kh_kw_{{r,k}}+w_k,
\end{equation}
where $k = 0,1,\hdots,M-1$, ${h}_k$ denotes the channel's transfer function (CTF), defined as ${h}_k = \sum_{n=0}^{L_h-1} \tilde{h}(n)e^{\frac{-j2\pi nk}{M}}$, while $w_k\sim \mathcal {N}(0, \sigma_w^{2})$, $w_{sd,k}\sim \mathcal {N}(0, \sigma_{sd}^{2})$, and $w_{r,k}\sim \mathcal {N}(0, \sigma _{r}^{2})$. Here, the signal-independent thermal noise is denoted by \( w_k \), while the ISDSN component is represented as \( \sqrt{|x_k h_k|} w_{{sd},k} \), and the RIN term is given by \( x_k h_k w_{r,k} \) \cite{yaseen2024signal}.}
The DCO-OFDM system model comprises $M$-parallel flat-fading systems. Hence, to recover the FD signal $\widehat{x}_k$ corresponding to the $k^{th}$ subcarrier, only single-tap equalization is needed at the receiver side. The transmitted bits are then recovered by the demapper/demodulator, which maps the complex symbols $\widehat{x}_k$ to the corresponding bits.
\subsection{ACO-OFDM system model}
The key distinction between this O-OFDM scheme and DCO-OFDM lies in the placement of data symbols. Specifically, in ACO-OFDM, consistent with the Hermitian symmetry condition $\tilde{x}_j = \tilde{x}^{\ast}_{M-j}, ~ \text{where} ~ M/2 +1 \leq j \leq M-1$, data symbols are assigned solely to the odd-indexed subcarriers within the first $M/2$ subcarriers. Consequently, only $M/4$ subcarriers convey information, in contrast to $M/2$ in DCO-OFDM. After the IFFT, the time-domain signal is real-valued and exhibits an anti-symmetric structure, wherein the first and second halves of the samples have identical amplitudes but opposite signs. This allows negative-valued samples to be clipped to zero without requiring a DC bias. The resulting waveform is then processed through the P/S, CP insertion, DAC, and LPF blocks to generate the continuous-time signal $x(t)$, which is subsequently passed to the optical modulator and transmitted over the frequency-selective multipath VLC channel \cite{saxena2023sparse}. At the receiver side, the principal operational difference from DCO-OFDM is the extraction of only the odd subcarriers after the FFT, which are then used for channel equalization and symbol demapping. The following section describes the multipath VLC CFR model.
\section{Multipath VLC CFR model}
\begin{figure}[t]
\centering
{\includegraphics[width=100mm, height=85mm]{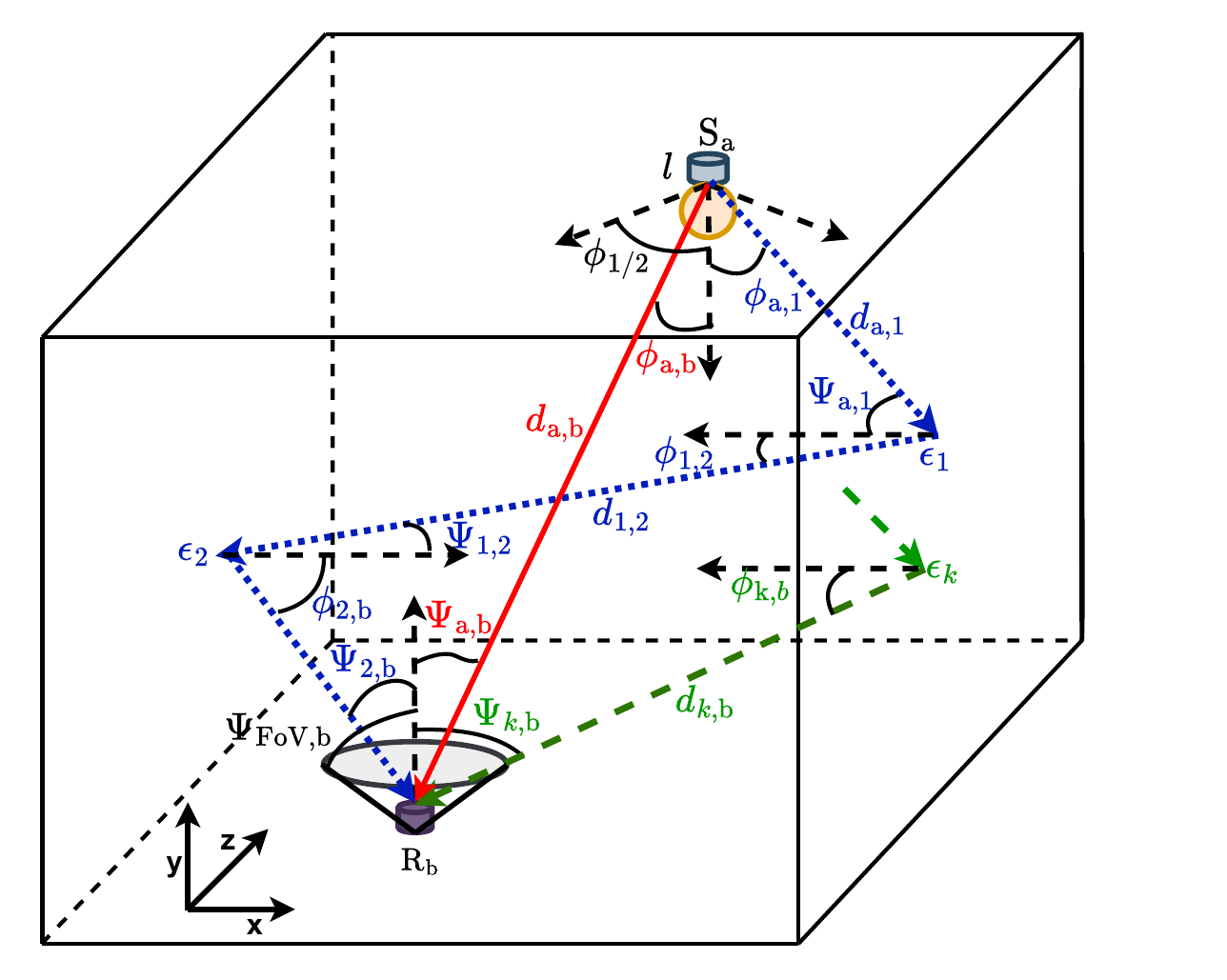}}
\caption{Arrangement of the receiver PD, reflectors, and transmitter LD in an indoor VLC environment. The red ``\textbf{---}" ray corresponds to the LoS path, the blue ``\textbf{$\cdot\cdot\cdot$}" component depicts the NLoS path, accounting for the $2^{\text{nd}}$ reflection, and the green ``\textbf{-- --}" ray shows the $\text{k}^{\text{th}}$ order multipath.}
\label{as1z_MMV}
\end{figure}

The multipath VLC channel model, encompassing both the NLoS and LoS elements, is established in this section. The light source employed is an LD, emitting radiation in accordance with a typical Lambertian structure $R_L(\phi)$ so that $\phi \in [-\pi/2, \pi/2 ] $ and featuring uniaxial symmetry. The quantity $R_L(\phi) = \frac{l+1}{2\pi}P_S\cos^l({\phi})$,
where the radiated power of the optical source is $P_S$, the particular radiating lobe's mode number is $l$, and $\phi$ denotes the irradiance angle of the source corresponding to the unit-length vector \textbf{\^{n}}$_\text{S}$ \cite{kahn1997wireless}. Here, the normal vector to the source radiating surface is given as \textbf{\^{n}}$_\text{S}$, and the mode number for the generalized Lambertian source is $l=1$. To achieve the highest radiance, the irradiance angle is set as $\phi = 0^{\circ}$, which results in $R_{L,\text{max}}(\phi) = \frac{l+1}{2\pi}P_S$. Furthermore, at half of $\text{R}_{L, \text{max}}$, we obtain the semi-angle of $ \phi_{1/2} = \text{arg}({R}_{L}(\phi) = {R}_{L,\text{max}}/2) = \cos^{-1}({2^{-1/l}}).$ The gain of a commonly used non-imaging concentrator, which amplifies the intensity of the signal received is represented as $\text{\textcrg} (\Psi) = \frac{\mu^{2}}{\sin ^{2} \Psi_{\text{FoV}}},$ if $0 \leq \Psi \leq \Psi_{\text{FoV}},$ and $0$ otherwise. To elaborate further, the optical concentrator's refractive index is $\mu$, $\Psi$ is in accordance with $\textbf{\^{n}}_\text{R}$ as the unit-length normal vector to the receiver's radiating surface, and the FoV angle obeys $\Psi_{\text{FoV}} \leq \pi/2$ \cite{kahn1997wireless}. Traditionally, the optical source ($\text{S}_\text{a}$) and the optical receiver ($\text{R}_\text{b}$) are represented as $\text{S}_\text{a}$ = $ \{ \textbf{r}_\text{S}, l, \textbf{\^{n}}_\text{S}\}$ and $\text{R}_\text{b}$ = $ \{ \textbf{r}_\text{R}, \Psi_{\text{FoV}}, \text{A}_\text{R}, \textbf{\^{n}}_\text{R} \} $, where $\textbf{r}_\text{R}$ is the optical receiver's position vector and $\textbf{r}_\text{S}$ is the optical source position vector. The receiver area is $\text{A}_\text{R}$. At the current state-of-the-art, the different reflecting surfaces (\textepsilon) like tiles, carpets, and walls are considered to have characteristics such as $ \{ \textbf{r}_\text{\textepsilon}, \pi/2, \text{A}_\text{\textepsilon}, \textbf{\^{n}}_\text{\textepsilon} \} $ and source characteristics as $ \{ \textbf{r}_\text{\textepsilon}, 1, \textbf{\^{n}}_\text{\textepsilon}\}$. The power radiated by \textepsilon~is $\text{\textrho}\text{P}_\text{\textepsilon}$, where $\text{P}_\text{\textepsilon}$ represents the incident power and $\text{\textrho}$ depicts the reflection coefficient of the reflecting surface, obeying $0 < \text{\textrho} < 1$ \cite{kahn1997wireless}. Fig. \ref{as1z_MMV} depicts the multipath VLC system with an optical source ($\text{S}_\text{a}$), two reflective elements ($\text{\textepsilon}_1, \text{\textepsilon}_2$), and the receiver ($\text{R}_\text{b}$) in an indoor environment. The LoS path in the multipath VLC channel emerging from the optical source $\text{S}_\text{a}$ = $ \{ \textbf{r}_\text{a}, l, \textbf{\^{n}}_\text{a}\}$ to the optical receiver $\text{R}_\text{b}$ = $ \{ \textbf{r}_\text{b}, \Psi_{\text{FoV}}, \text{A}_{\text{R}_\text{b}}, \textbf{\^{n}}_\text{b} \} $ is modeled by employing a Dirac delta function with scaled and delayed versions formulated as \cite{kahn1997wireless}
\begin{align} 
h^{(0)}(t; \text{R}_\text{b}, \text{S}_\text{a}) & = 
\frac{l+1}{2 \pi d_{\text{a}, \text{b}}^{2}} \text{A}_{\text{R}_\text{b}} \cos ^{l} (\phi_{\text{a}, \text{b}}) \cos  (\Psi_{\text{a}, \text{b}})\text{\textcrg} (\Psi_{\text{a}, \text{b}}) \text{rect}\left( \frac{\Psi_{\text{a},\text{b}}}{\Psi_{\text{FoV},\text{b}}}\right) \delta\left(t-\frac{d_{\text{a}, \text{b}}}{c}\right) 
\nonumber \\ 
& = h_{\text{LoS}} \delta\left(t-\frac{d_{\text{a}, \text{b}}}{c}\right), 
\end{align}
where the incidence angle at $\text{R}_\text{b}$ is $\Psi_{\text{a}, \text{b}}$, the area of the PD is  $\text{A}_{\text{R}_\text{b}}$, speed of light is $c$, $d_{\text{a}, \text{b}}$ is the distance between $\text{R}_\text{b}$ and $\text{S}_\text{a}$, the FoV of $\text{R}_\text{b}$ is $\Psi_{\text{FoV},\text{b}}$, $\delta(.)$ is the Dirac delta function, the irradiance from $\text{S}_\text{a}$ is represented by $\phi_{\text{a}, \text{b}}$, and the mode number corresponding to the $\text{S}_\text{a}$ is $l$. Furthermore, $d_{\text{a}, \text{b}} = ||\textbf{r}_\text{b} - \textbf{r}_\text{a}||=\sqrt{R^2 + L^2}$, $\cos (\Psi_{\text{a}, \text{b}}) = \textbf{\^{n}}_\text{b}.(\textbf{r}_\text{b} - \textbf{r}_\text{a})/d_{\text{a}, \text{b}} = L/d_{\text{a}, \text{b}}$, $\cos (\phi_{\text{a}, \text{b}}) = \textbf{\^{n}}_\text{a}.(\textbf{r}_\text{b} - \textbf{r}_\text{a})/d_{\text{a}, \text{b}} =L/d_{\text{a}, \text{b}}$, where $L$ represents the vertical separation between the LD and the receiver plane, the user is assumed to be positioned within a circular region of radius \( R \). Additionally, the gain of the LoS signal exhibits a monotonically decreasing behavior as a function of $d_{\text{a}, \text{b}}$, and it is formulated as
\begin{align}
    h_{\text{LoS}} = \frac{l+1}{2 \pi d_{\text{a}, \text{b}}^{2}} \text{A}_{\text{R}_\text{b}} \cos ^{l} (\phi_{\text{a}, \text{b}}) \cos  (\Psi_{\text{a}, \text{b}})\text{\textcrg} (\Psi_{\text{a}, \text{b}})
\text{rect}\left( \frac{\Psi_{\text{a},\text{b}}}{\Psi_{\text{FoV},\text{b}}}\right). 
\end{align}
Thus, the CFR of the LoS link can be expressed as \cite{schulze2016frequency}
\begin{align}\label{p1}
    h_{\text{LoS},f} = h_{\text{LoS}} \exp (-j2\pi f \tau_{\text{LoS}}),
\end{align}
\hspace{-1.2mm}where $\tau_{\text{LoS}} = d_{\text{a}, \text{b}}/c$.
Similarly, the NLoS component between $\text{S}_{\text{a}}$ and $\text{R}_{\text{b}}$ corresponding to the ${m}^{th}$ reflection and ${M}_{r}$ reflective surfaces is given by
\begin{align}
&h_{\text{a} \rightarrow b}^{(m)}\left(t ; \text{R}_{\text{b}}, \text{S}_{\text{a}}\right)
=\frac{l+1}{2 \pi} \sum_{i=1}^{{M}_{r}} \frac{\text{A}_i \rho_{i}}{d_{\text{a}, i}^{2}} \cos ^{l} (\phi_{\text{a}, i}) \cos (\psi_{\text{a}, i}) \text{rect}\left(\frac{\psi_{\text{a}, i}}{\pi/2}\right) h^{(m-1)}\left(t-\frac{d_{\text{a}, i}}{c} ; \text{R}_{\text{b}}, \text{\textepsilon}_{i}\right).
\end{align}
Furthermore, the authors of \cite{schulze2016frequency,jungnickel2002physical} proposed a heuristic channel model for the NLoS component that departs from conventional microscopic approaches by disregarding individual reflections and instead modeling the overall channel behavior. This simplified formulation assumes spatial uniformity and yields an analytical expression that closely aligns with experimental results \cite{schulze2016frequency,jungnickel2002physical}. Inspired by sphere analysis, the model replaces the sphere’s surface and volume with the room parameters $A_{\text{room}}$ and $V_{\text{room}}$, respectively. The NLoS gain is expressed as \cite{schulze2016frequency,jungnickel2002physical}
\begin{align}
   \eta_{\text{NLoS}} = \frac{A_{\text{R}_{\text{b}}}}{A_{\text{room}}} \frac{\rho}{1 - \rho}, 
\end{align}
which holds for uniform reflectivity $\rho$, derived from the geometric series
\begin{align}
 \lim_{\ell \to \infty} \sum_{\ell=1}^{L} \rho^\ell = \frac{\rho}{1 - \rho}.   
\end{align}
Thus, the impulse response for the NLoS component is modeled as \cite{schulze2016frequency,jungnickel2002physical}
\begin{align}
    h_{\text{NLoS}}(t) = \frac{\eta_{\text{NLoS}}}{\tau_{\text{NLoS}}} \exp (-t/\tau_{\text{NLoS}}) u(t),
\end{align}
where $u(t)$ is the unit step function with a time constant
\begin{align}
    \tau_{\text{NLoS}} = \frac{-4V_{\text{room}}}{c\ln \rho A_{\text{room}}}.
\end{align}
Thus, the CFR for the NLoS component is given as \cite{schulze2016frequency,jungnickel2002physical}
\begin{align}
    h_{\text{NLoS},f} = \frac{\eta_{\text{NLoS}}}{1 + j \frac{f}{f_0} }, \text{~with~} f_0 = \frac{1}{2\pi \tau_{\text{NLoS}}},
\end{align}
where $f_0$ is the cutoff frequency. The total transfer function for the LoS and NLoS components is given as 
\begin{align}
    h_{f} = h_{\text{LoS},f}+h_{\text{NLoS},f}.
\end{align}
Upon denoting the frequency spacing by $\Delta f$, the computation of $h_{f}$ is restricted to frequencies $f = k\Delta f$ for $k = 0, 1, \ldots, M/2 - 1$, due to the Hermitian symmetry property $h_{-f} = h_{f}^*$, which arises from the real-valued nature of the TD CIR. Additionally, the parameter $f_0$ is defined as $f_0 = M \Delta f $, where $f_0/2$ is the simulation bandwidth \cite{schulze2024dispersive}. Prior to executing the IFFT, a raised cosine window is employed to smoothen the spectral data, thereby minimizing artifacts in the resulting TD response. Thus, the discrete CFR for the $k^{th}$ subcarrier is given by 
\begin{align} \label{p2}
    h_{k} &= h_{\text{LoS},k}+h_{\text{NLoS},k} = h_{\text{LoS}} \exp (-j2\pi k\Delta f \tau_{\text{LoS}}) + \frac{\eta_{\text{NLoS}}}{1 + j \frac{k}{M}} \nonumber \\ 
    & = \ h_{\text{LoS}} \exp \left(j\theta_{1,k}\right) + |h_{\text{NLoS},k}|\exp({j\theta_{2,k}}),
\end{align}
where $|h_{\text{NLoS},k}| = \frac{\eta_{\text{NLoS}}}{\sqrt{1 + \left(\frac{k}{M}\right)^2}}$, $\theta_{1,k} = -2\pi k\Delta f \tau_{\text{LoS}}$, and $\theta_{2,k} = \tan^{-1}{\left(\frac{-k}{M}\right)}$ \cite{schulze2016frequency,schulze2024dispersive}. 
It is assumed that the receiver's position is uniformly distributed over a circular region with a maximum cell radius $R$, and the corresponding PDF of the radial distance $r$ is given by \cite{jani2019performance} 
\begin{align}
    f_{r}(r) = \frac{2r}{R^2}.
\end{align}
Employing the receiver's PDF in conjunction with ($\ref{p1}$), and upon appropriate mathematical derivations, the PDF corresponding to the LoS CFR $h_{\text{LoS,k}}$ is given as \cite{yaseen2023channel}

\begin{align} 
f_{h_{\text{LoS,k}}}(h_{\text{LoS,k}}) = {\begin{cases}\Upsilon h_{\mathrm{LoS,k}}^{\frac{-2}{l+3}-1}& \hspace{-3mm} h_{\mathrm{LoS,k,min}} \leq h_{\mathrm{LoS,k}}\leq h_{\mathrm{LoS,k,max}}, \\ 0 & \hspace{-3mm} \text{otherwise}, \end{cases}} 
\end{align}
where $\Upsilon$ and $\varphi$ are given as \cite{mitra2018minimum} 
\begin{align} 
\Upsilon =\frac{{2}\varphi^{\bigl(\frac{2}{l+3}\bigr)}((l+1)L^{(l+1)})^{\frac{2}{l+3}}}{ (l+3)R^{2}},  
 \end{align}
\begin{align}
   \varphi = \frac{\text{\textcrg} (\Psi_{\text{a}, \text{b}})\text{rect} \left(\frac{\Psi_{\text{a},\text{b}}}{\Psi_{\text{FoV},\text{b}}}\right)A_{\text{R}_{\text{b}}}\eta}{2\pi}. 
\end{align}
The corresponding values of $h_{\text{LoS},\mathrm{k,min}}$ and $h_{\text{LoS},\mathrm{k,max}}$ are given as \cite{mitra2018minimum}
\vspace{-2mm}
\begin{align} 
h_{\text{LoS},\mathrm{k,min}}&=\frac{\varphi(l+1)L^{l+1}}{\left(R^{2}+L^{2}\right)^{\frac{l+3}{2}}}\exp (j\theta_{1,k}), 
\end{align}
\vspace{-2mm}
\begin{align}
h_{\text{LoS},\mathrm{k,max}}&=\frac{\varphi(l+1)L^{l+1}}{R^{l+3}}\exp (j\theta_{1,k}). 
\end{align} 
Following appropriate mathematical derivations from (\ref{p2}), the expression for $h_{\text{LoS},k}$ can be reformulated as
\vspace{-2mm}
\begin{align}
    h_{\text{LoS},k} = h_k - g_k,
\end{align}
where $g_k = h_{\text{NLoS},k}$. Thus, by a suitable transformation, the PDF for $h_k$ is expressed as
\begin{align} \label{p5}
    f_{ h_k}(h_k) &= f_{h_{\text{LoS},k}}(h_k-g_k) = {\begin{cases}\Upsilon (h_k-g_k)^{-\frac{2}{l+3}-1}& h_{k,\mathrm{min}} \leq h_{k}\leq h_{k,\mathrm{max}}, \\ 0 & \text{otherwise}, \end{cases}} 
\end{align}
where we have:
\vspace{-3mm}
\begin{align}
  h_{k,\mathrm{min}} = \frac{\varphi(l+1)L^{\left(l+1\right)}}{(R^2+L^2)^\frac{l+3}{2}}\exp (j\theta_{1,k}) + g_k,   
\end{align}
\vspace{-3mm}
\begin{align}
   h_{k,\mathrm{max}} = \frac{\varphi(l+1)L^{\left(l+1\right)}}{R^{\left(l+3\right)}}\exp (j\theta_{1,k}) + g_k.  
\end{align}
The mean value of $h_k$ is given as 
\begin{align}
    \mu_{h_{k}} = \frac{(l+3)\Upsilon \left((h_{k,\mathrm{max}}-g_k)^{\frac{l+1}{l+3}}-(h_{k,\mathrm{min}}-g_k)^{\frac{l+1}{l+3}}\right)}{l+1} + g_k.
\end{align}
\section{Bayesian Cramér-Rao Lower Bound}
In this section, the BCRLB is derived in the presence of ISDSN, RIN, and thermal noise for a random VLC channel. In estimation theory and statistics, the BCRLB pertains to the estimation of deterministic, yet unknown, parameters. It establishes that the precision of any unbiased estimator is fundamentally limited by the Fisher information, wherein the inverse of the Fisher information serves as a lower bound on the variance of such estimators. An unbiased estimator that attains this bound is considered efficient, indicating that it achieves the minimum possible MSE among all unbiased estimators \cite{yaseen2023channel,saxena2023sparse}.

Let us now consider a scenario where a sequence of \( N \) frames of O-OFDM pilot symbols is transmitted over the $k^{th}$ subcarrier, in order to facilitate the estimation of the CFR, \( h_k \), prior to actual data transmission. Upon representing the O-OFDM pilot symbols as \( \boldsymbol{x}_k = [x_k(1), x_k(2), \dots, x_k(N)]^T \) and the received signal vector of the $k^{th}$ subcarrier as $ \boldsymbol{y}_k = [y_k(1) , $ $ y_k(2), \dots, y_k(N)]^T $, we arrive at \cite{yaseen2024signal,saxena2023sparse}
\begin{align} 
\boldsymbol{y}_k=h_k\boldsymbol{x}_k+\sqrt{h_k\mathrm{diag}(\boldsymbol{x}_k)}\boldsymbol{w}_{sd,k}+ h_k\mathrm{diag}(\boldsymbol{x}_k)\boldsymbol{w}_{r,k} + \boldsymbol{w}_k,
\end{align}
where we have \( \boldsymbol{w}_k = [w_k(1), w_k(2), \dots, w_k(N)]^T,~\boldsymbol{w}_{sd,k} = [w_{sd,k}(1), w_{sd,k}(2), \dots,w_{sd,k}(N)]^T \), and \( \boldsymbol{w}_{r,k} = [w_{r,k}(1), w_{r,k}(2), \dots, w_{r,k}(N)]^T \). The components of \( \boldsymbol{w}_k \) follow an independent and identically distributed (IIDs) Gaussian distribution, expressed as \( \boldsymbol{w}_k \sim \mathcal{N}(0, \sigma_w^2 \mathbf{I}_N) \). Similarly, the noise terms \( \boldsymbol{w}_{sd,k} \) and \( \boldsymbol{w}_{r,k} \) are modeled as \( \boldsymbol{w}_{sd,k} \sim \mathcal{N}(0, \sigma_{sd}^2 \mathbf{I}_N) \) and \( \boldsymbol{w}_{r,k} \sim \mathcal{N}(0, \sigma_{r}^2\mathbf{I}_N) \), respectively. 

Considering the $k^{th}$ subcarrier, the prior CFR PDF \( f_{h_k}(h_k) \) at the receiver, the Bayesian information is expressed as
\begin{align} \label{p8}
J_k=J_{h,k} + J_{y,k}. 
\end{align}
Here, the Bayesian Fisher information for the channel prior is \( J_{h,k}\), and that for the observation model is \( J_{y,k} \), where \( J_{h,k}\) can be formulated as
\begin{align} \label{p9}
J_{h,k}&=\mathbb{E}\left\lbrace \left(\frac{\partial }{\partial h_k}\ln f_{h_k}(h_k)\right)^{2}\right\rbrace = \mathbb{E}\left\lbrace \left( \left(\frac{-2}{l+3}-1\right)\frac{1}{h_k-g_k}\right)^{2}\right\rbrace \nonumber\\ &=\left(\frac{2}{l+3}+1\right)^{2}{\mathbb{E}} \left\lbrace \frac{1}{(h_k-g_k)^2}\right\rbrace = -\Upsilon \frac{(l+5)^{2}}{2(l+3)(l+4)} \bigg (h_{k,\mathrm{max}}^{\left(\frac{-2l-8}{l+3}\right)}-h_{k,\mathrm{min}}^{\left(\frac{-2l-8}{l+3}\right)}\bigg).
\end{align}
\begin{figure*}[ht]
\small
\begin{flalign} \label{p3}
&\frac{\partial }{\partial h_k}\ln f(\boldsymbol{y}_k|h_k)  \nonumber \\
&= \frac{1}{2}\sum _{n=1}^{N}\left[\frac{x_k(n)[2y_k(n) -\sigma_{{sd}} ^{2} - 2h_k\sigma^2_{{r}}x_k(n)-2h_kx_k(n)]}{\Sigma_{h,k}^{(n)}} +\frac{(\sigma_{{sd}} ^{2}x_{k}(n) + 2h_k\sigma^2_{{r}}x^2_k(n))(y_{k}(n)-h_kx_{k}(n))^{2}}{\left(\Sigma_{h,k}^{(n)}\right)^2}\right]. 
\end{flalign} \vspace{-6mm}
\begin{flalign} \label{p4}
&\frac{\partial^2}{\partial h^2_k} \ln f(\boldsymbol{y}_k \mid h_k)  = 
 -\frac{1}{2} \sum_{n=1}^{N} \Biggl[
\frac{2x_k^2(n)(\sigma_{r}^2+1) }{\Sigma_{h,k}^{(n)}}- \frac{\left( \sigma_{{sd}}^2 x_k(n) + 2h_k \sigma_{r}^2 x_k^2(n) \right)^2}{\left(\Sigma_{h,k}^{(n)}\right)^2} - \frac{(y_k(n) - h_k x_k(n))^2 2\sigma_{r}^2 x_k^2(n)
}{\left( \Sigma_{h,k}^{(n)} \right)^2}\nonumber\\&
+ \frac{4x_k(n) (y_k(n) - h_k x_k(n))(\sigma_{sd}^2 x_k(n) + 2h_k \sigma_{r}^2 x_k^2(n))}{\left( \Sigma_{h,k}^{(n)} \right)^2} + \frac{2 \left( \sigma_{{sd}}^2 x_k(n) + 2h_k \sigma_{r}^2 x_k^2(n) \right)^2 (y_k(n) - h_k x_k(n))^2}{\left( \Sigma_{h,k}^{(n)}\right)^3} 
\Biggr].
\end{flalign}\vspace{-6mm}
\begin{flalign} \label{p7}
J_{y,k} =(l + 3) \Biggl[(h_k-g_k)^{\frac{-2}{l + 3}}\Biggl\{ \varphi_1 \Biggl(\frac{\Gamma_1}{g_k-h_1} -\frac{\Gamma_2}{g_k-h_2}\Biggr) +\varphi_2\Gamma_3 +\varphi_3\Gamma_4 + \varphi_4\Biggl(\frac{ \Gamma_1 }{g_k-h_1}-\frac{\Gamma_2}{g_k-h_2}  \Biggr) \Biggr\} \Biggr]_{h_{k,\min}}^{h_{k,\max}}. &&
\end{flalign}\vspace{-4mm}
\begin{flalign} \label{p10}
 J_{y,k}&=   \Biggl[(h_k-g_k)^{\frac{-2}{l + 3}} \Biggl\{-\frac{Np^2\Upsilon(l+3)}{2(\sigma^2_{w}+g_k\sigma^2_{{sd}}p)}{}_2F_1\left(\frac{-2}{l + 3}, 1; \frac{l + 1}{l + 3}; \frac{-p (h_k-g_k)\sigma^2_{{{sd}}}}{\sigma^2_{w}+\sigma^2_{{sd}}g_kp}\right) \nonumber \\ 
 & - \frac{N p^4\sigma^4_{{sd}}\Upsilon(l + 3)}{4( \sigma_w^2+g_k\sigma^2_{{sd}}p)^2} 
{}_2F_1\left(\frac{-2}{l + 3}, 2; \frac{l + 1}{l + 3}; \frac{-p (h_k-g_k)\sigma^2_{{{sd}}}}{\sigma^2_{w}+\sigma^2_{{sd}}g_kp}\right)\Biggl\}\Biggl]_{h_{k,\min}}^{h_{k,\max}}. &&
\end{flalign}\vspace{-4mm}
\begin{flalign} \label{p11}
    &J_{y,k} = \frac{(l+3)Np\Upsilon }{2g_k}\nonumber \\ 
    & \times \left[(h_k-g_k)^\frac{-2}{l+3}\left\{\frac{-1}{\sigma^2_{sd}} {}_2F_1\left(\frac{-2}{l + 3}, 1; \frac{l + 1}{l + 3}; \frac{- (h_k-g_k)}{g_k}\right)-\frac{p}{2g_k} {}_2F_1\left(\frac{-2}{l + 3}, 2; \frac{l + 1}{l + 3}; \frac{- (h_k-g_k)}{g_k}\right) \right\}\right]_{h_{k,\min}}^{h_{k,\max}}. &&
\end{flalign}\vspace{-6mm}
\begin{flalign} \label{p12}
    J_{y,k} &= \frac {-Np^2\Upsilon (l+3)}{g^2_k} \left[(h_k-g_k)^\frac{-2}{l+3} {}_2F_1\left(\frac{-2}{l + 3}, 2; \frac{l + 1}{l + 3}; \frac{ -(h_k-g_k)}{g_k}\right)\right]_{h_{k,\min}}^{h_{k,\max}}. &&
\end{flalign}
\hrulefill
\end{figure*}
Furthermore, $J_{y,k}$ may be expressed as
\vspace{-2mm}
\begin{align} 
J_{y,k}=-\mathbb{E}\left\lbrace \frac{\partial ^{2}}{\partial h^{2}_k}\ln f(\boldsymbol{y}_k|h_k)\right\rbrace. 
\end{align}
Assuming that all the \(N\) samples are IIDs, the likelihood function \(f(\boldsymbol{y}_k|h_k)\) of \(h_k\) is expressed as follows
\begin{align}
f(\boldsymbol{y}_k|h_k)= \prod _{n=1}^{N}\frac{1}{\sqrt{2\pi \Sigma_{h,k}^{(n)} }}\exp \biggl({-\frac{1}{2}\sum _{n=1}^{N}\frac{(y_{k}(n)-h_kx_{k}(n))^{2}}{\Sigma_{h,k}^{(n)}}}\biggr),
\end{align}
where $\Sigma_{h,k}^{(n)} = \sigma_w^2 + \sigma_{{sd}}^2 h_k x_k(n) + \sigma_{r}^2 h_k^2 x_k^2(n)$.
%
%
By applying the natural logarithm to \( f(\boldsymbol{y}_k|h_k) \) and taking the partial derivative with respect to $h_k$, we arrive at (\ref{p3}).
Subsequently, the second-order partial derivative is formulated in (\ref{p4}). Accordingly, by evaluating the expectation terms \( \mathbb{E}\{y_k(n)\} = h_kx_k(n) \), \( \mathbb{E}\{(y_k(n) - h_kx_k(n))^2\} = \Sigma_{h,k}^{(n)}\), the Fisher information may be expressed as 
\begin{align} \label{p26}
J_{y,k} = \frac{1}{2} \mathbb{E} \left\lbrace \sum_{n=1}^{N} \Bigg[\frac{2x_k^2(n)}{\Sigma_{h,k}^{(n)}} + \frac{\left( \sigma_{{sd}}^2 x_k(n) + 2h_k \sigma_{r}^2 x_k^2(n) \right)^2}{\left( \Sigma_{h,k}^{(n)}\right)^2}\Bigg]\right\rbrace.
\end{align}
Furthermore, assuming \( x_k(n) = p \) for all \( n = 1, \dots, N \), the Fisher information can be expressed as
\begin{align} \label{p6}
J_{y,k} = Np^{2} \mathbb{E} \left\lbrace \frac{1}{\Sigma_{p,k}}\right\rbrace +\frac{Np^2}{2}\mathbb{E} \left\lbrace \frac{(\sigma_{{sd}}^2 + 2h_k \sigma_{r}^2 p)^2}{( \Sigma_{p,k})^2 }\right\rbrace, 
\end{align}
where we have $\Sigma_{p,k} = \sigma_w^2 + \sigma_{{sd}}^2 h_kp + \sigma_{r}^2 h_k^2 p^2$. Moreover, by leveraging (\ref{p5}), (\ref{p6}) may be restated as follows
\begin{align}
J_{y,k} & = Np^{2}\Upsilon \int _{h_{k,\mathrm{min}}}^{h_{k,\mathrm{max}}}\frac{(h_k-g_k)^{\frac{-(l+5)}{(l+3)}}}{\Sigma_{p,k}}\,dh_k  +\frac{Np^2}{2}\Upsilon \int _{h_{k,\mathrm{min}}}^{h_{k,\mathrm{max}}}\frac{(h_k-g_k)^{\frac{-(l+5)}{(l+3)}}(\sigma_{{sd}}^2 + 2h_k \sigma_{r}^2 p)^2}{(\Sigma_{p,k})^2}\,dh_k. 
\end{align}
Following appropriate mathematical simplifications, the above expression is reformulated as (\ref{p7}), 
where $\varphi_1 =-\frac{\Upsilon N}{2\sigma^2_{{r}}(h_1-h_2)}$, $\varphi_2=-\frac{\Upsilon N(2h_1\sigma^2_{{r}}p^2+\sigma^2_{{sd}}p)^2}{4\sigma _{{r}}^{4}p^4(g_k-h_1)^2\left(h_1 - h_2\right)^2}$, $\varphi_3=-\frac{\Upsilon N(2h_2\sigma^2_{{r}}p^2+\sigma^2_{{sd}}p)^2}{4\sigma _{{r}}^{4}p^4(g_k-h_2)^2\left(h_1 - h_2\right)^2}$, $\varphi_4=\frac{\Upsilon N(2h_1\sigma^2_{{r}}p^2+\sigma^2_{{sd}}p)(2h_2\sigma^2_{{r}}p^2+\sigma^2_{{sd}}p)}{2(h_1-h_2)^3\sigma^4_{{r}}p^4}, h_{1}=\frac{-\sigma^2_{{sd}} - \sqrt{\sigma^4_{{sd}}-4\sigma^2_{w}\sigma^2_{{r}}}}{{2\sigma^2_{{r}}p}}$, $h_{2}=\frac{-\sigma^2_{{sd}} + \sqrt{\sigma^4_{{sd}}-4\sigma^2_{w}\sigma^2_{{r}}}}{{2\sigma^2_{{r}}p}}$, $\Gamma_1= {}_2F_{1}\Bigl(\frac{-2}{(l + 3)},1; $ $\frac{(l + 1)}{(l + 3)}; \frac{-(h_k-g_k)}{g_k-h_1}\Bigr),$ $\Gamma_2= {}_2F_{1}\left(\frac{-2}{(l + 3)},1; \frac{(l + 1)}{(l + 3)}; \frac{-(h_k-g_k)}{g_k-h_2}\right)$, $\Gamma_3= {}_2F_{1}\Bigl(\frac{-2}{(l + 3)},2; \frac{(l + 1)}{(l + 3)}; \frac{-(h_k-g_k)}{g_k-h_1}\Bigr),$ and $\Gamma_4= {}_2F_{1}\Bigl(\frac{-2}{(l + 3)},2; \frac{(l + 1)}{(l + 3)}; \frac{-(h_k-g_k)}{g_k-h_2}\Bigr)$.
Thus, the BCRLB for the $k^{th}$ subcarrier is expressed as 
\begin{align}
    \sigma^2_{k,\epsilon} \geq \frac{1}{J_k} ,
\end{align}
where $\sigma^2_{k,\epsilon}=\mathbb{E}\{[\widehat{h}_k - h_k]^2\}$ is the MSE and \( J_k \) is derived by substituting equations (\ref{p7}) and (\ref{p9}) into (\ref{p8}). Next, we discuss the Fisher information under special cases of RIN, ISDSN, and thermal noise.

\textit{\textbf{Remark $1$: Fisher Information in absence of ISDSN and RIN}}:
Setting \( \sigma^2_{{sd}} = 0 \) and \( \sigma^2_{r} = 0 \) in (\ref{p7}) yields the Fisher information function as follows
\begin{align} 
J_{y,k}=\sum _{n=1}^{N}\frac{x_{k}^{2}(n)}{\sigma ^{2}_{w}}. 
\end{align}
Thus, under the assumptions of \( x_i = p \) and in the absence of ISDSN and RIN, the expected Fisher information becomes
\vspace{-2mm}
\begin{align} \label{p18}
J_{y,k} = \frac{N p^{2}}{\sigma ^{2}_{w}}. 
\end{align}
Thereby, the Bayesian information may be given as follows
\begin{align}
J_k = \frac{Np^{2}}{\sigma ^{2}_{w}} + J_{h,k}.
\end{align}

\begin{figure*}[!htb]
\normalsize
\begin{flalign} \label{p13}
    J_{y,k} = Np^{2}\Upsilon \int _{h_{k,\mathrm{min}}}^{h_{k,\mathrm{max}}}\frac{(h_k-g_k)^{\frac{-(l+5)}{(l+3)}}}{\sigma_w^2  + \sigma_{r}^2 h^2_kp^2}\,dh_k +2Np^{2}\Upsilon \int _{h_{k,\mathrm{min}}}^{h_{k,\mathrm{max}}}\frac{(h_k-g_k)^{\frac{-(l+5)}{(l+3)}}( h_k \sigma_{r}^2 p^2)^2}{(\sigma_w^2  + \sigma_{r}^2 h^2_k p^2)^2}\,dh_k. &&
\end{flalign} \vspace{-4mm}
\begin{flalign} \label{p14}
    J_{y,k} = - \left(\frac{1}{2\sigma^2_{{r}}}+\frac{p^2}{4}\right)\frac{N \Upsilon (l + 3)}{g^2_k} \left[{(h_k-g_k)^\frac{-2}{l+3}}{}_2F_1\left(\frac{-2}{l + 3}, 2; \frac{l + 1}{l + 3}; \frac{ -(h_k-g_k)}{g_k}\right)\right]_{h_{k,\min}}^{h_{k,\max}}. &&
\end{flalign} \vspace{-4mm}
\begin{flalign} \label{p15}
    J_{y,k} = 
 \frac{-Np^2\Upsilon (l+3)}{4g^2_k} \left[(h_k-g_k)^{\frac{-2}{l+3}}{}_2F_1\left(\frac{-2}{l + 3}, 2; \frac{l + 1}{l + 3}; \frac{-(h_k-g_k)}{g_k}\right)\right]_{h_{k,\min}}^{h_{k,\max}}. &&
\end{flalign} \vspace{-4mm}
\begin{flalign} \label{p16}
&J_{y,k} = Np(l+3)\Upsilon \Biggl[(h_k-g_k)^\frac{-2}{l+3} \Bigg\{\frac{\sigma^2_{{r}}}{2\sigma^2_{{sd}}(\sigma^2_{{sd}}+\sigma^2_{{r}}pg_k)}{}_2F_1\left(\frac{-2}{l + 3}, 1; \frac{l + 1}{l + 3}; \frac{ -(h_k-g_k)p\sigma^2_{{r}}}{(\sigma^2_{{sd}}+\sigma^2_{{r}}pg_k)}\right)\nonumber \\
&+ \frac{1}{2\sigma^2_{{sd}}g_k}{}_2F_1\left(\frac{-2}{l + 3}, 1; \frac{l + 1}{l + 3}; \frac{ -(h_k-g_k)}{g_k}\right) -\frac{p}{4(l+3)g^2_k}{}_2F_1\left(\frac{-2}{l + 3}, 2; \frac{l + 1}{l + 3}; \frac{-(h_k-g_k)}{g_k}\right) \nonumber \\
&-\frac{p^3\sigma^4_{{r}}}{2(\sigma^2_{{sd}}+\sigma^2_{{r}}g_kp)^2}{}_2F_1\left(\frac{-2}{l + 3}, 2; \frac{l + 1}{l + 3}; \frac{-(h_k-g_k)p\sigma^2_{{r}}}{(\sigma^2_{{sd}}+\sigma^2_{{r}}g_kp)}\right) -\frac{ p\sigma^2_{{r}}}{\sigma^2_{sd}g_k}{}_2F_1\left(\frac{-2}{l + 3}, 1; \frac{l + 1}{l + 3}; \frac{-(h_k-g_k)}{g_k}\right) \nonumber \\
& +\frac{ p^3\sigma^4_{{r}}}{2\sigma^2_{{sd}}(\sigma^2_{{sd}}+\sigma^2_{{r}} g_kp)}{}_2F_1\left(\frac{-2}{l + 3}, 1; \frac{l + 1}{l + 3}; \frac{-p\sigma^2_{{r}}(h_k-g_k)}{\sigma^2_{{sd}}+\sigma^2_{{r}}g_kp}\right) \Bigg\}\Biggl]_{h_{k,\min}}^{h_{k,\max}}. &&
\end{flalign} \vspace{-4mm}
\begin{flalign} \label{p17}
J_{y,k} & = \frac{N(l+3)\Upsilon }{g^2_k}\Bigg[(h_k-g_k)^{\frac{-2}{l+3}}\Bigg\{\frac{-1}{2\sigma^2_{{r}}} {}_2F_1\left(\frac{-2}{l + 3}, 2; \frac{l + 1}{l + 3}; \frac{ -(h_k-g_k)}{g_k}\right) \nonumber \\
& - p^2
{}_2F_1\left(\frac{-2}{l + 3}, 2; \frac{l + 1}{l + 3}; \frac{ -(h_k-g_k)}{g_k}\right) \Bigg\} \Bigg]_{h_{k,\min}}^{h_{k,\max}}. &&
\end{flalign}
\hrulefill
\end{figure*}
\textit{\textbf{Remark $2$: Fisher Information for special cases of RIN}}:
In this section, we examine the Fisher information for special cases of RIN.
\begin{enumerate}
\item In the case of a negligible level of RIN, i.e., $\sigma _{r}^{2} \to 0$, the Fisher information reduces to the expression given in (\ref{p10}). Additionally, at high power $p$ associated with $\sigma _{r}^{2} \to 0$, the Fisher information can be approximately represented in (\ref{p11}).
\item Conversely, at a high RIN level, the Fisher information can be represented as determined in (\ref{p12}).

\end{enumerate}

\textit{\textbf{Remark $3$: Fisher Information for special cases of ISDSN}}:
This section describes the Fisher information results for two special cases of ISDSN values.
\begin{enumerate}
\item In the case of a negligible level of ISDSN, i.e., $\sigma _{sd}^{2} \to 0$, the Fisher information reduces to (\ref{p13}). In the absence of the ISDSN, the Fisher information at high power $p$ can be generally represented by (\ref{p14}).

\item Conversely, at a high ISDSN level, the Fisher information can be represented as (\ref{p15})
\end{enumerate}

\textit{\textbf{Remark $4$: Fisher Information for special cases of thermal noise}}:
Here, we consider Fisher information under two specific thermal noise scenarios

\begin{enumerate}
\item In the case of a negligible level of thermal noise, i.e., $\sigma _{w}^{2} \to 0$, the Fisher information reduces to (\ref{p16}). Assuming thermal noise $\sigma_w^2$ is small and power $p$ is high, the Fisher information is given by (\ref{p17}).

\item On the other hand, when thermal noise $\sigma_w^2$ is high, the Fisher information is represented as
\begin{align}
J_{y,k} =0.
\end{align}
\end{enumerate}

\textit{\textbf{Remark $5$: Fisher Information for special cases of RIN, ISDSN, and thermal noise}}:
\begin{enumerate}
    \item In the presence of ISDSN and thermal noise \( \boldsymbol{w} \), the quantity \( J_{y,k} \) increases proportionally with the order of \( p \), since the numerator's dependence on \( p \) dominates the denominator, as indicated in (\ref{p10}). Consequently, the BCRLB exhibits an inverse relationship at high power levels. Conversely, in the absence of RIN and ISDSN, the BCRLB decreases proportionally to \( \frac{1}{p^2} \), as established in (\ref{p18}).  
    \item According to (\ref{p12}), under conditions of high RIN and transmitted power, \( J_{h,k} \) is primarily influenced by the signal power and remains unaffected by other noise sources.
    \item Furthermore, in the presence of thermal noise alone, a significant reduction in the BCRLB is observed with an increase in the transmitted power.
\end{enumerate}
\section{ISDSN and RIN-based VLC CE techniques}
In this section, the characteristics and performance of five estimators, LS, LMMSE, ML, MAP, and MMSE, proposed for the estimation of CFR in O-OFDM VLC systems are examined in detail.
\subsection{Least square (LS) estimator}
For the LS estimator, the estimated \( \widehat{h}_{k,\text{LS}} \) is given as
\begin{align} \label{p25}
\widehat h_{k,\mathrm{LS}}= \frac {\boldsymbol {x}^{H}_k\boldsymbol{y}_k}{|| \boldsymbol{x}_k ||^{2}}.  
\end{align}
The efficiency of the LS estimator is assessed through the MSE, which corresponds to the variance associated with the estimation error. Accordingly, the estimation error is given by
\begin{align} 
\epsilon _{k,\mathrm { LS}}=\widehat h_{k,\mathrm { LS}}-h_k.  
\end{align}
It is straightforward to verify that the LS estimator is unbiased, i.e., $\mathbb{E}\{\epsilon_{k,\mathrm{LS}}\} = 0$. Moreover, the expression for $\epsilon_{k,\mathrm{LS}}$ can be formulated as follows
\begin{align} \label{p19}
\epsilon _{k,\mathrm { LS}}= \frac {\boldsymbol{x}^{H}_k}{||\boldsymbol {x}_k||^{2}}\left ({{\sqrt {h_k{\mathrm { diag}}(\boldsymbol{x_k})}\boldsymbol{w}_{{sd},k}+{h_k\mathrm { diag}}(\boldsymbol{x}_k)\boldsymbol{w}_{r,k}+\boldsymbol {w}_k}}\right)\cdot  
\end{align}
From (\ref{p19}), it is apparent that \( \epsilon_{k,\mathrm { LS}} \) follows a Gaussian distribution, i.e., \( \epsilon_{k,\mathrm { LS}} \sim \mathcal{N}(0, \sigma^2_{\epsilon_{k,\mathrm { LS}}}) \). The corresponding variance \( \sigma^2_{\epsilon_{k,\mathrm { LS}}} \), which also represents the MSE, is given by
\begin{align}
\sigma ^{2}_{\epsilon _{k,\mathrm { LS}}}=\frac {\sum _{n=1}^{N}x_{k}^{2}(n)(\sigma ^{2}_{w}+\sigma ^{2}_{sd}h_kx_{k}(n)+\sigma ^{2}_{r}h_k^{2}x^{2}_{k}(n))}{\left ({{\sum _{n=1}^{N}x_{k}^{2}(n)}}\right)^{2}}\cdot  
\end{align}
For \( x_{k}(n) = p\), the expression for \(\sigma^2_{\epsilon_{k,\mathrm{LS}}}\) admits a further simplification as
\begin{align} \label{p20}
\sigma ^{2}_{\epsilon _{k,\mathrm { LS}}}=\frac {\sigma ^{2}_{w}+\sigma ^{2}_{sd}h_kp+\sigma ^{2}_{r}h_k^{2}p^2}{Np^2}\cdot  
\end{align}
Observe from (\ref{p20}), that an increase in power causes $\sigma ^{2}_{\epsilon _{k,\mathrm { LS}}}$ to reach a saturation point, whereas an increase in the number of pilots effectively reduces the error variance $\sigma ^{2}_{\epsilon _{k,\mathrm { LS}}}$, thereby improving the performance of the LS estimator.
\begin{figure*}[t]
\normalsize
\begin{flalign} \label{p219}
     &\widehat{h}_{k,\mathrm{LMMSE}}=\frac{\sigma ^{2}_{k}\Vert \boldsymbol{x}_k\Vert ^{2}\boldsymbol{x}_k^{H}\left(\boldsymbol{y}_k-\boldsymbol{x}_k\mu_{k}\right)}{\sigma ^{2}_{k}\Vert \boldsymbol{x}_k\Vert ^{4}+\mu _{k}\sigma ^{2}_{sd}\boldsymbol{x}_k^{H}\mathrm{diag}(\boldsymbol{x}_{k})\boldsymbol{x}_k+ (\sigma^2_{k}+\mu^2_{k})\sigma^2_{{r}}\boldsymbol{x}_k^{H}\mathrm{diag}^2(\boldsymbol{x}_{k})\boldsymbol{x}_k+\sigma ^{2}_{w}\Vert \boldsymbol{x}_k\Vert ^{2}}
     +\mu_{k}. &&
\end{flalign} \vspace{-4mm} 
\begin{flalign} \label{p22}
\sigma ^{2}_{\epsilon _{k,\mathrm{LMMSE}}}=\sigma ^{2}_{k}\left[ 
\frac{
\mu _{k}\sigma ^{2}_{sd}\boldsymbol{x}^{H}_k
\mathrm{diag}(\boldsymbol{x_k})\boldsymbol{x_k} 
+(\sigma^2_{k}+\mu^2_{k})\sigma^2_{{r}}\boldsymbol{x}^{H}_k
(\mathrm{diag}(\boldsymbol{x_k}))^2\boldsymbol{x_k}
+\sigma ^{2}_{w}\Vert \boldsymbol{x_k}\Vert ^{2}
}{
\sigma ^{2}_{k}\Vert \boldsymbol{x_k}\Vert ^{4}
+\mu _{k}\sigma ^{2}_{sd}\boldsymbol{x}^{H}_k
\mathrm{diag}(\boldsymbol{x_k})\boldsymbol{x_k}+(\sigma^2_{k}+\mu^2_{k})\sigma^2_{{r}}\boldsymbol{x}^{H}_k
(\mathrm{diag}(\boldsymbol{x_k}))^2\boldsymbol{x_k}
+\sigma ^{2}_{w}\Vert \boldsymbol{x_k}\Vert ^{2}
}
\right]. &&
\end{flalign}
\hrulefill
\end{figure*}

\subsection{Linear minimum mean square error (LMMSE) estimator}
The LMMSE estimator is a linear function of the observation $y_k$ that minimizes the MSE \( \mathbb{E}\{[\widehat{h}_k - h_k]^2\} \). Thus, the estimated channel \( \widehat{h}_{k,\text{LMMSE}} \) is derived as  (\ref{p219}). Given the relationship \( \widehat{h}_{k,\text{LMMSE}} = h_k + \varepsilon_{k,\text{LMMSE}} \), the estimation error \( \varepsilon_{k,\text{LMMSE}} \) follows a Gaussian distribution, \( \varepsilon_{k,\text{LMMSE}} \sim \mathcal{N}(0, \sigma^2_{\varepsilon_{k,\text{LMMSE}}}) \). The variance \( \sigma^2_{\varepsilon_{k,\text{LMMSE}}} \) is then formulated as outlined in (\ref{p22}). Assuming \( x_k(n) = p \) for all \( n = 1, \dots, N \), and observing that \( \boldsymbol{x}^T_k \text{diag}(\boldsymbol{x}_k) \boldsymbol{x}_k = \sum_{i=1}^{N} (x_k(n))^3 \), (\ref{p22}) can be succinctly expressed as follows:
\begin{align} \label{p23}
 \sigma^2_{\varepsilon_{k,\text{LMMSE}}} = \frac{\sigma^2_k[\sigma^2_{sd}\mu_kp+(\sigma^2_k+\mu_k^2)\sigma^2_rp^2+\sigma^2_w]}{\sigma^2_kp^2N+\mu_k\sigma^2_{{sd}}p+(\sigma^2_k+\mu_k^2)\sigma^2_rp^2+\sigma^2_w}. 
\end{align}

\textit{\textbf{Remark $6$}}: From (\ref{p23}), the following observations are made: 
i) As \( p \to 0 \), the MSE approaches \( \sigma_k^2 \), dominated by thermal noise, with the estimate relying solely on prior statistics.  
ii) As the transmitted power increases, the MSE converges to \( \frac{\sigma_{r}^2(\sigma^2_k+\mu_k^2)\sigma^2_k}{\sigma^2_{k}N+(\sigma^2_k+\mu_k^2)\sigma^2_{{r}}} \), where it is governed by the prior mean \( \mu_k^2 \), by \( \sigma^2_k \), and by the number of pilots \( N \).

\subsection{Maximum likelihood (ML) estimator}
The ML estimator determines the value of \( h_k \) that maximizes the likelihood function. Accordingly, the channel gain in the VLC system is obtained by equating the derivative of the log-likelihood given by (\ref{p3}) to zero and solving for \( h_k \).
Upon assuming identical pilot symbols across all transmissions, i.e., \( x_k(n) = p \) for every \( i \), the corresponding estimate of \( h_k \) is obtained as follows
\begin{align} \mathcal{A}_k\widehat{h}_k^{3}+\mathcal{B}_k\widehat{h}_k^{2}+\mathcal{C}_k\widehat{h}_k+\mathcal{D}_k=0,  
\end{align}
where the coefficients $\mathcal{A}_k, \mathcal{B}_k, \mathcal{C}_k,$ and $\mathcal{D}_k,$ are defined as follows
\begin{align}
 \mathcal{A}_k& =2Np^{4}\sigma _{r}^{4},  \\  \vspace{-2mm}
 \mathcal{B}_k& =3N\sigma _{sd}^{2}\sigma _{r}^{2}p^{3}+N\sigma _{sd}^{2}p^{3}+2p^{3}\sigma _{r}^{2}\sum _{n=1}^{N}y_{k}(n),  \\ \vspace{-2mm}
 \mathcal{C}_k& =N\sigma _{sd}^{4}p^{2}+2N\sigma _{r}^{2}\sigma _{w}^{2}p^{2}+2Np^{2}\sigma _{w}^{2}-2\sigma _{r}^{2}p^{2}\sum _{n=1}^{N}y_{k}^{2}(n),
\\ \vspace{-2mm} 
\mathcal{D}_k& =N\sigma _{sd}^{2}p\sigma _{w}^{2}-2p\sigma _{w}^{2}\sum _{n=1}^{N}y_{k}(n)-\sigma _{sd}^{2}p\sum _{n=1}^{N}y_{k}^{2}(n).  
\end{align}
\begin{figure*}[t]
\small
\begin{flalign} \label{p24}
&\frac{1}{2}\sum _{n=1}^{N} \Biggl[ \frac{(\sigma_{{sd}} ^{2}x_{k}(n) + 2h_k\sigma^2_{{r}}x^2_k(n))(y_{k}(n)-h_kx_{k}(n))^{2}}{(\Sigma_{h,k}^{(n)})^{2}} +\frac{2x_{k}(n)(y_{k}(n) - h_kx_{k}(n)-\sigma_{{sd}} ^{2} - 2h_k\sigma^2_{{r}}x_k(n))}{\Sigma_{h,k}^{(n)}} \Biggr] \nonumber \\
& -\frac{(l+5)}{(l+3)}\frac{1}{h_k-g_k} =0. &&
\end{flalign} 
\hrulefill
\vspace{-3mm}
\end{figure*}
Due to the analytical intractability of determining the MSE associated with the ML estimator, its performance is evaluated heuristically through simulations. 

\textit{\textbf{Remark $7$:}} Observe from (\ref{p24}) that when ISDSN and RIN are not present, i.e, $\sigma^2_{{sd}} = \sigma^2_{{r}} = 0$, then
\begin{align} 
\frac{\partial }{\partial h_k}\ln f(\boldsymbol{y}_k|h_k) =& \sum _{n=1}^{N} \frac{x_{k}(n)(y_{k}(n)- h_kx_{k}(n))}{\Sigma_{h,k}^{(n)}}. 
\end{align}
Assuming \( x_k(n) = p \) for all \( n = 1, \ldots, N \), the ML estimate of the channel gain is given by \( \widehat{h}_{k,\text{ML}} = \frac{1}{Np} \sum_{i=1}^N y_k(n) \), which corresponds to the LS estimate in (\ref{p25}). This indicates that, when ISDSN and RIN are absent, the ML and LS estimators yield identical performance. As a result, the MSE for both methods becomes \( \frac{\sigma_w^2}{Np^2} \).

\textit{\textbf{Remark $8$:}} Although the proposed ML estimator exhibits bias, i.e., \( \mathbb{E}\{\widehat{h}_{k,\mathrm{ML}}\} \neq \mathbb{E}\{h_k\} \), it becomes asymptotically unbiased as \( N \to \infty \). That is, for sufficiently large symbol lengths, \( \mathbb{E}\{\hat{h}_{k,ML}\} \to \mathbb{E}\{h_k\} \). Consequently, under large-sample conditions, the ML estimator becomes unbiased and attains the Fisher information, i.e., \(\widehat{h}_{k,\mathrm{ML}} \sim \mathcal{N}(h_k, 1/J_{y,k}) \) \cite{yaseen2023channel}, where $J_{y,k}$ corresponds to the Fisher information matrix as defined in (\ref{p26}). Under the assumption of \( x_k(n) = p ~(\forall n = 1,\dots,N) \), the expression for $J_{y,k}$ simplifies to
\begin{align}
J_{y,k}= \frac {Np^{2}(2\Sigma_{p,k}+(\sigma _{sd}^{2}+2h_kp\sigma _{r}^{2})^2)}{2\Sigma_{p,k}^2}.
\end{align}

\subsection{Maximum \textit{a posterior} probability (MAP) estimator}
The MAP estimator derives \( h_k\) by selecting the value that maximizes this posterior, resulting in the following estimate
\begin{align} \label{p27}
\widehat{h}_{k,\mathrm{MAP}} = \arg \max _{h_k} f(h_k|\boldsymbol{y}_k) = \arg \max _{h_k} \ln f(h_k|\boldsymbol{y}_k),     
\end{align}
where we have \( f(h_k|\boldsymbol{y}_k) = \frac{f(\boldsymbol{y}_k|h_k) f_{h_k}(h_k)}{f_{\boldsymbol{y}_k}(\boldsymbol{y}_k)} \). From equation (\ref{p27}), it is clear that the maximization of the log-posterior function \( \ln f(h_k|\boldsymbol{y}_k) \) enables the estimation of \( \widehat{h}_{k,\mathrm{MAP}} \), which can be determined by solving the following equation
\begin{align} \label{p28}
&\frac{\partial }{\partial h_k}\ln\biggl( \frac{f(\boldsymbol{y}_k|h_k) f_{h}(h_k)}{f_{\boldsymbol{y}_k}(\boldsymbol{y}_k)}\biggr)=0 \nonumber\\
&\Rightarrow 
\frac{\partial }{\partial h_k}\left[ -\frac{1}{2}\sum _{n=1}^{N}\ln \left(\Sigma_{h,k}^{(n)}\right) -\frac{1}{2}\sum _{n=1}^{N}\frac{(y_{i}-h_kx_{i})^{2}}{\Sigma_{h,k}^{(n)}} \right.\nonumber\\ 
& \left. +\ln\Upsilon-\frac{N}{2}\ln \left(2\pi\right)-\frac{(l+5)}{(l+3)}\ln h_k-\ln f_{\boldsymbol{y}_k}(\boldsymbol{y}_k)\right]=0,
\end{align}
where $f_{\boldsymbol{y}_k}(\boldsymbol{y}_k)$ is independent of $h_k$. Upon simplifying the expression in (\ref{p28}), we arrive at (\ref{p24}).
Assuming a constant pilot power, i.e., \( x_k(n) = p \) for all \( n = 1, \ldots, N \), the VLC channel gain estimate \( \widehat{h}_k \) is determined by solving the equation 
\begin{align}
A_k\widehat{h}^{4}_k+B_k\widehat{h}^{3}_k+C_k\widehat{h}^{2}_k+D_k\widehat{h}_k+E_k=0,
\end{align}
where the coefficients \( A_k, B_k, C_k, D_k, \) and \( E_k \) are defined at the top of the next page.
The numerical approach requires the VLC channel gain \( h_k \) to be a real and positive quantity. Deriving a closed-form representation for this solution remains analytically intractable. Consequently, the performance of the MAP estimator is assessed through Monte Carlo-based simulations.
\begin{figure*}[t]
\small
\begin{flalign}
A_k& =-2Np^{4}\sigma _{r}^{4}-2\left(\frac{l+5}{l+3}\right)\sigma_{r}^4p^4. &&\\  \vspace{-2mm}
B_k& =-3N\sigma _{sd}^{2}\sigma _{r}^{2}p^{3}-N\sigma _{sd}^{2}p^{3}+2p^{3}\sigma _{r}^{2}\sum _{n=1}^{N}y_{k}(n)-4\left(\frac{l+5}{l+3}\right)\sigma _{r}^{2}\sigma _{sd}^{2}p^3
+2g_kN\sigma^4_{r}p^4. &&\\ \vspace{-2mm}
C_k& =-N\sigma _{sd}^{4}p^{2}-2N\sigma _{r}^{2}\sigma _{w}^{2}p^{2}-2Np^{2}\sigma _{w}^{2}+2\sigma _{r}^{2}p^{2}\sum _{n=1}^{N}y_{k}^{2}(n)-4\left(\frac{l+5}{l+3}\right)\sigma _{r}^{2}\sigma _{w}^{2}p^2-2\left(\frac{l+5}{l+3}\right)\sigma^4_{sd}p^2 \nonumber \\
&+3g_kN\sigma^2_{r}\sigma^2_{sd}p^3 +g_kN\sigma^2_{sd}p^3+2g_kp^3\sigma^2_{r}\sum _{n=1}^{N}y_k(n). && \\  \vspace{-4mm}
D_k& =-N\sigma _{sd}^{2}p\sigma _{w}^{2}+2p\sigma _{w}^{2}\sum _{n=1}^{N}y_{k}(n)+\sigma _{sd}^{2}p\sum _{n=1}^{N}y_{k}^{2}(n)-4\left(\frac{l+5}{l+3}\right)\sigma _{r}^{2}\sigma _{w}^{2}p
+g_kN\sigma^4_{sd}p^2 + 2g_kN\sigma^2_{r}\sigma^2_{w} \nonumber &&\\ \vspace{-4mm}
&+ 2g_kNp^2\sigma^2_{w} - 2g_k\sigma^2_{r}p^2\sum_{n=1}^{N}y_{k}^2(n). && \\ \vspace{-2mm}
E_k& =-2\left(\frac{l+5}{l+3}\right)\sigma _{r}^{4}+g_kN\sigma^2_{sd}\sigma^2_{w}p -2g_kp\sum_{i=1}^{N}y_k(n)\sigma^2_{w} -g_kp\sigma^2_{sd}\sum_{i=1}^{N}y^2_k(n). && 
\end{flalign} \vspace{-2mm}
\hrulefill
\vspace{-3mm}
\end{figure*}

\subsection{Minimum mean square error (MMSE) estimator}
The MMSE technique seeks to reduce the MSE, thereby offering a statistically optimal solution. It determines the channel gain by evaluating the conditional expectation over the posterior distribution \( f(h_k|\boldsymbol{y}_k) \), leading to the estimator expression \( \widehat{h}_{k,\mathrm{MMSE}} = \mathbb{E}\left\lbrace h_k|\boldsymbol{y}_k \right\rbrace \). Upon utilizing Bayes’ theorem, this estimator can be alternatively represented as in (\ref{p29}).
Assuming that \( x_k(n) = p \) holds for every \( n = 1, \dots, N \), the formulation of \( \widehat{h}_{k,\text{MMSE}} \) simplifies to
\begin{align} \label{p31}
\widehat{ h}_{k,\mathrm{MMSE}}=& \frac{\Upsilon }{f_{\boldsymbol{y}_k}(\boldsymbol{y}_k)}\int _{h_{k,\mathrm{min}}}^{h_{k,\mathrm{max}}} h_k(h_k-g_k)^{\frac{-(l+5)}{l+3}} \prod _{n=1}^{N}\frac{1}{\sqrt{2\pi\Sigma_{p,k}}}\times \exp \left({\sum _{n=1}^{N}\frac{-(y_k(n)-h_kp)^{2}}{2\Sigma_{p,k}}}\right)dh_k.
\end{align}
Moreover, \( {f_{\boldsymbol{y}_k}(\boldsymbol{y}_k)} \) represents the marginal PDF of \( \boldsymbol{y}_k \), formulated as seen in (\ref{p30}).
Assuming \( x_k(n) = p \) for all \( n = 1, \ldots, N \), (\ref{p30}) simplifies to
\begin{align} \label{p32}
f_{\boldsymbol{y}_k}(\boldsymbol{y}_k) = \frac{\Upsilon}{(2\pi )^{\frac{N}{2}}} \int _{h_\mathrm{k,min}}^{h_\mathrm{k,max}}\frac{(h_k-g_k)^{{\frac{-(l+5)}{l+3}}}}{({\Sigma_{p,k}})^\frac{N}{2}}\exp \left(\sum_{n=1}^{N}\frac{-(y_{k}(n)-h_kp)^{2}}{2\Sigma_{p,k}}\right)dh_k. 
\end{align}
Accordingly, based on (\ref{p31}) and (\ref{p32}), the MMSE estimate of the channel gain, \( \widehat{h}_{k,\text{MMSE}} \), is derived in (\ref{l1}).
\begin{figure*}[!t]
\small
\begin{flalign} \label{p29}
\widehat{h}_\mathrm{k,MMSE}&= \int _{h_\mathrm{k,min}}^{h_\mathrm{k,max}} \frac{h_k f(\boldsymbol{y}_k|h_k)f_{h_k}(h_k)}{f_{\boldsymbol{y}_k}(\boldsymbol{y}_k)} dh_k \nonumber \\
&= \frac{\Upsilon }{f_{\boldsymbol{y}_k}(\boldsymbol{y}_k)} \int _{h_\mathrm{k,min}}^{h_\mathrm{k,max}} h_k(h_k-g_k)^{\frac{-(l+5)}{(l+3)}} \left(\prod _{n=1}^{N}\frac{1}{\sqrt{2\pi\Sigma_{h,k}^{(n)}}} \exp \left({\sum _{n=1}^{N}\frac{-(y_{k}(n)-h_kx_{k}(n))^{2}}{2\Sigma_{h,k}^{(n)}}}\right)\right)dh_k. &&
\end{flalign}  \vspace{-4mm}
\begin{flalign} \label{p30}
{f_{\boldsymbol{y}_k}(\boldsymbol{y}_k)} &= \int _{h_\mathrm{k,min}}^{h_{ \mathrm{k,max}}}\left(\prod _{n=1}^{N}f(y_{k}(n)|h_k)\right)f_{h_k}(h_k)dh_k \nonumber\\  \vspace{-2mm}
&=  \frac{\Upsilon}{(2\pi)^{N/2}} \int _{h_\mathrm{k,min}}^{h_\mathrm{k,max}}(h_k-g_k)^{{\frac{-(l+5)}{l+3}}}\left(\prod _{n=1}^{N}\frac{1}{\sqrt{\Sigma_{h,k}^{(n)}}} \exp \left(\frac{-(y_{k}(n)-h_kx_{k}(n))^{2}}{2\Sigma_{h,k}^{(n)}}\right)\right)dh_k. &&
\end{flalign}  \vspace{-2mm}
\begin{flalign} \label{l1}
\widehat{h}_{\mathrm{k,MMSE}} =\frac{\int_{h_{\mathrm{k,min}}}^{h_{\mathrm{k,max}}}
\frac{h_k(h_k-g_k)^{\frac{-(l+5)}{l+3}}}{\left( \Sigma_{p,k} \right)^{\frac{N}{2}}}
\exp\left(\sum_{n=1}^{N}
\frac{-(y_k(n) - h_k p)^2}{2\Sigma_{p,k}}\right)dh_k}{
\int_{h_{\mathrm{k,min}}}^{h_{\mathrm{k,max}}}
\frac{(h_k-g_k)^{\frac{-(l+5)}{l+3}}}{(\Sigma_{p,k})^\frac{N}{2}}\exp\left( \sum_{n=1}^{N}\frac{-(y_k(n) - h_k p)^2}{
2 \Sigma_{p,k} }\right)dh_k}. &&
\end{flalign}
\hrulefill
\end{figure*}
\begin{figure*}[t]
	\centering
    \captionsetup[subfigure]{justification=centering}
	\subfloat[]{\label{bv1}\includegraphics[width=55mm,height=50mm]{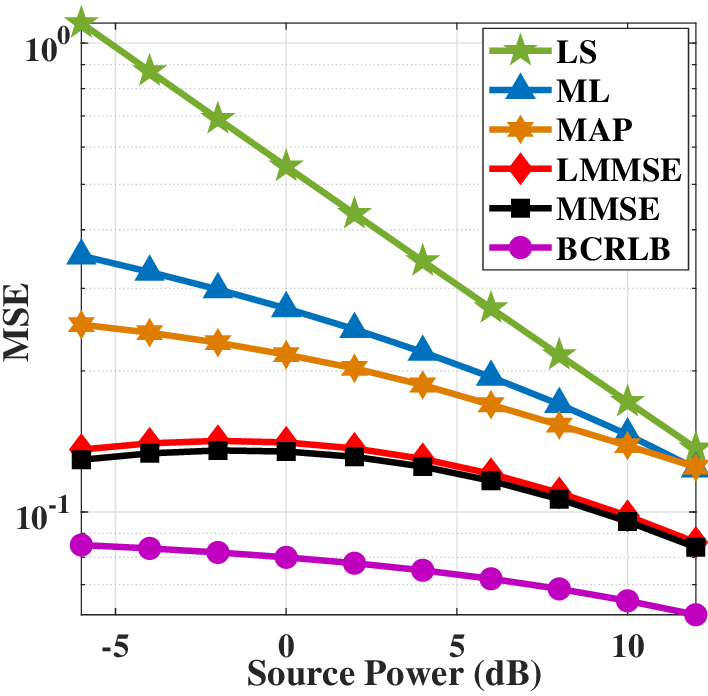}}
	\subfloat[]{\label{sh2}\includegraphics[width=55mm,height=50mm]{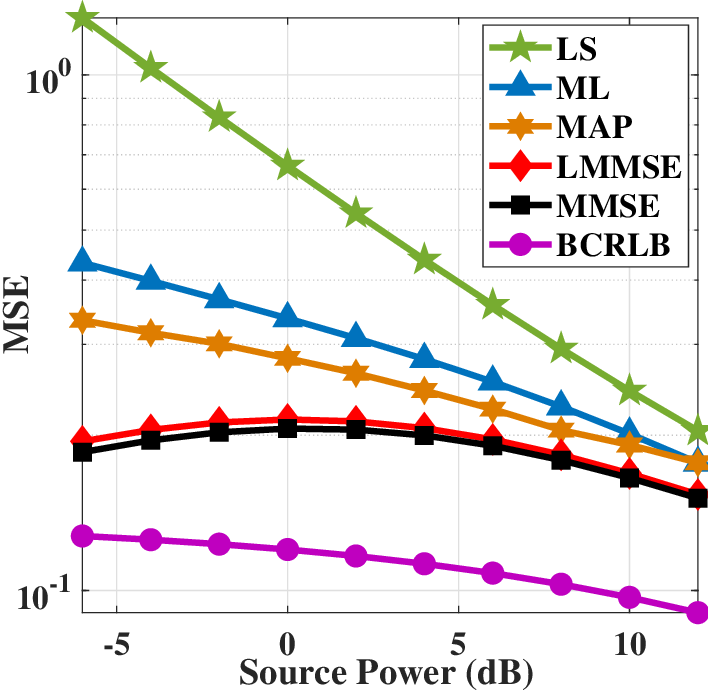}} 
    \subfloat[]    {\label{sh11}\includegraphics[width=55mm,height=50mm]{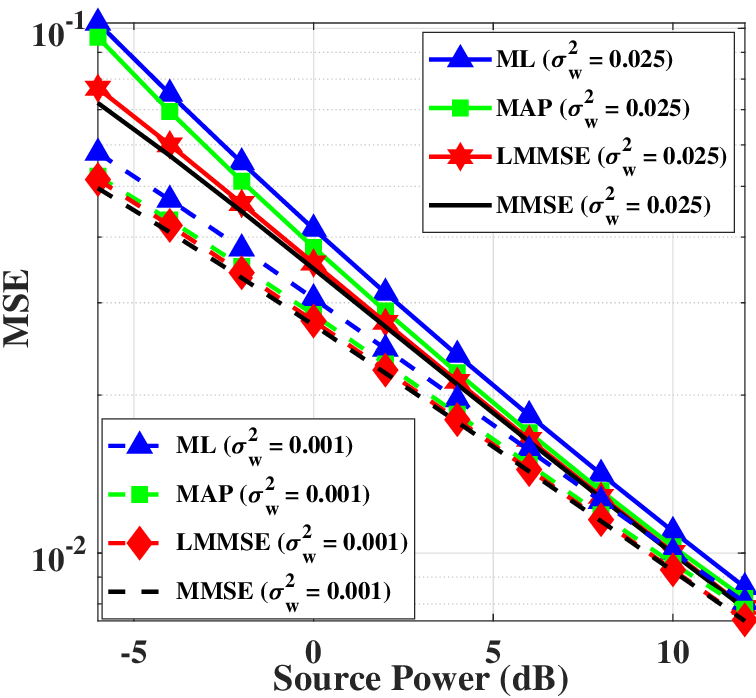}}
	\caption{The MSE versus source power is shown for the SISO ISDSN-RIN VLC system for: (a) proposed estimators with BCRLB for DCO-OFDM system; (b) \textcolor{black}{proposed estimators with BCRLB for ACO-OFDM system;} (c) different values of \( \sigma_w^2 \) for DCO-OFDM system.}
	\label{R1}
\end{figure*}
\begin{figure*}[t]
	\centering
    \captionsetup[subfigure]{justification=centering}
	\subfloat[]{\label{sh21}\includegraphics[width=55mm,height=50mm]{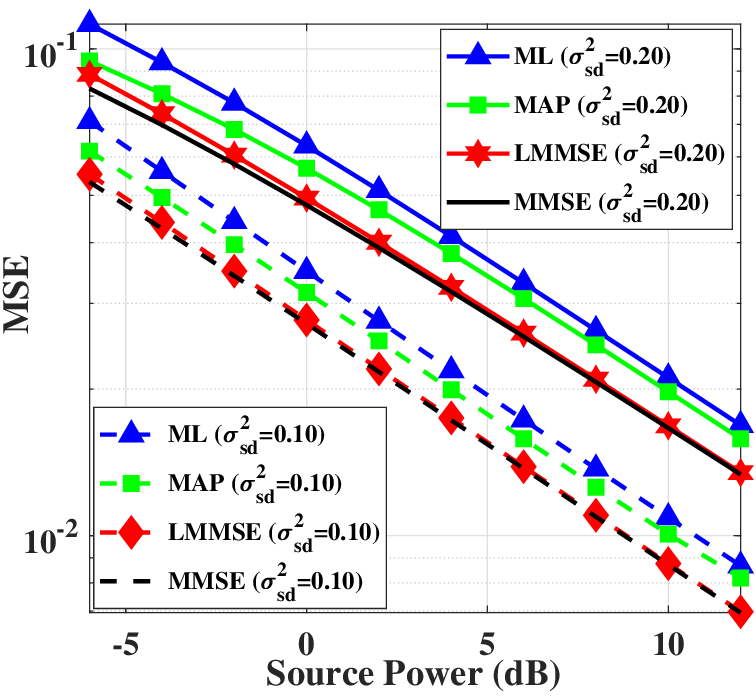}}
    \subfloat[]{\label{sh3}\includegraphics[width=55mm,height=50mm]{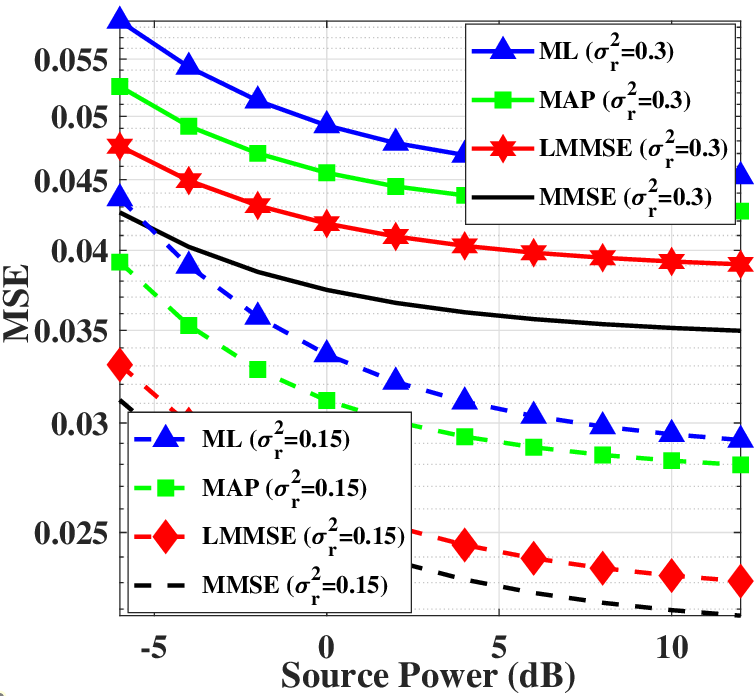}} 
    \subfloat[]{\label{sh4}\includegraphics[width=55mm,height=50mm]{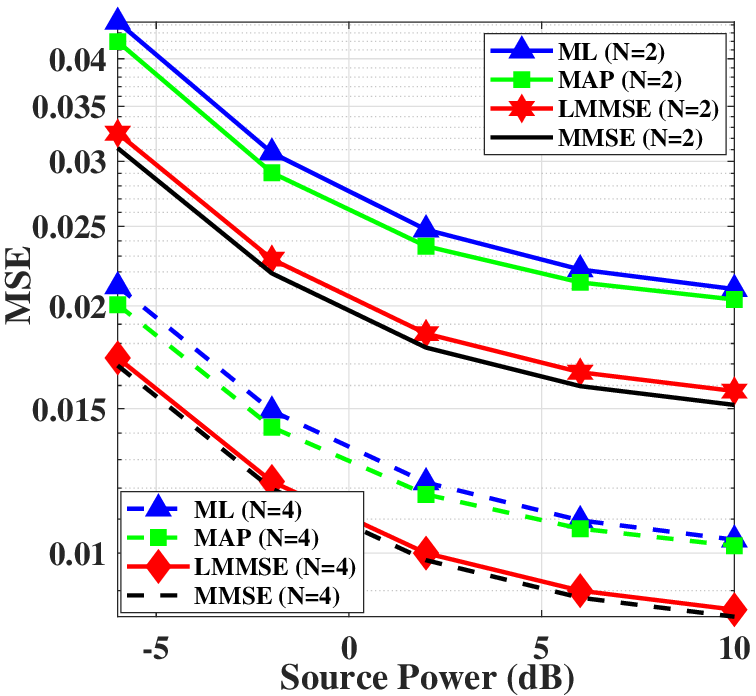}}
	\caption{The MSE versus source power is shown for the SISO DCO-OFDM ISDSN-RIN VLC system under: (a) different values of \( \sigma^2_{sd}\); (b) different values of \( \sigma^2_{r}\); (c) different numbers
    of frames.}
	\label{R1w}
\end{figure*}

\section{Simulation Results}
A comparative evaluation of the proposed estimators is conducted for the SISO O-OFDM VLC system incorporating ISDSN and RIN, under a statistically varying channel model. The simulation setup employs the following parameters: $A_{\text{R}_{\text{b}}} = 1 \text{~cm}^2$, $\eta = 1$ Amp/W, $\text{\textcrg} (\Psi_{\text{a,b}}) = 1$, $l = 1$, $R = 0.5$ m, $M = 1024$, $B_\text{DC} = 7$ dB, and $L = 2$ m \cite{yaseen2024signal,saxena2023sparse}.

Fig. \ref{R1}\subref{bv1} illustrates the MSE performance of the proposed MMSE-based estimator alongside alternative approaches for the DCO-OFDM ISDSN RIN VLC system, evaluated under fixed noise variances $\sigma_w^2 = 0.01$, $\sigma_{sd}^2 = 1$, $N = 1$, and $\sigma_r^2 = 0.01$. As evident from the figure, the LS estimator exhibits the highest MSE among the methods compared, whereas the MMSE estimator achieves the lowest. Estimators that utilize prior statistical knowledge consistently deliver improved estimation accuracy relative to their non-Bayesian counterparts. \textcolor{black}{Additionally, the MSE results are benchmarked against the BCRLB, with the MMSE-based scheme exhibiting a performance closely aligned with this bound. The MSE performance of the various estimation strategies employed in the ACO-OFDM system is presented in Fig. \ref{R1}\subref{sh2}. A similar trend is observed, wherein the MMSE-based estimator consistently achieves the best overall performance. These observations collectively confirm the superior efficacy of the proposed MMSE method under the combined influence of ISDSN, RIN, and thermal noise in O-OFDM VLC systems.}

Fig. \ref{R1}\subref{sh11} illustrates the effect of thermal noise on the performance of different estimators in the DCO-OFDM ISDSN RIN VLC framework, with $\sigma^2_{sd} = 0.1$, $N = 1$, and $\sigma_r^2 = 0.001$ held constant. A noticeable MSE increases is observed for all schemes at $\sigma_w^2 = 0.025$ in comparison to $\sigma_w^2 = 0.001$, with the MMSE estimator consistently yielding the lowest MSE throughout. This highlights the adverse influence of increasing $\sigma_w^2$ on estimation precision. Moreover, the performance gap between the estimators becomes more significant as $\sigma_w^2$ rises. However, under high power and reduced thermal noise levels, the MSE values of all estimators tend to converge. These findings collectively reinforce the efficiency of the proposed MMSE method in delivering more accurate estimates under diverse thermal noise conditions.

Fig. \ref{R1w}\subref{sh21} characterizes the impact of ISDSN noise on the performance of various estimators for the DCO-OFDM ISDSN RIN VLC system, with fixed parameters of $\sigma^2_{w} = 0.005$, $N = 1$,  and $\sigma_r^2 = 0.001$. It is evident that all estimators exhibit increased MSE at $\sigma^2_{sd} = 0.2$ in comparison to $\sigma^2_{sd} = 0.1$, with the MMSE estimator consistently achieving the lowest MSE among the methods evaluated. These observations indicate that the MMSE approach maintains superior performance across different ISDSN levels, while the presence of ISDSN significantly impairs the overall estimation accuracy.

\begin{figure*}[t]
	\centering
    \captionsetup[subfigure]{justification=centering}
	\subfloat[]{\label{p21}\includegraphics[width=65mm,height=55mm]{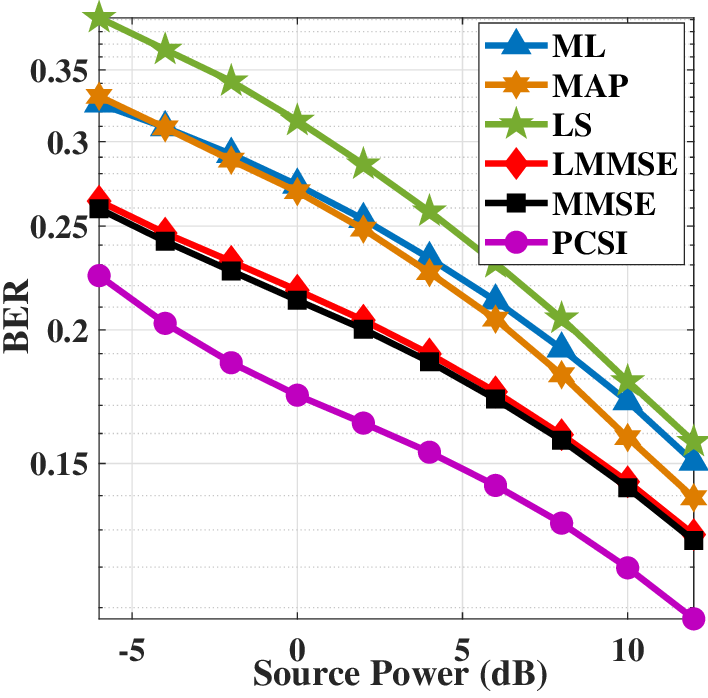}} \hspace{10mm}
	\subfloat[]{\label{p231}\includegraphics[width=65mm,height=55mm]{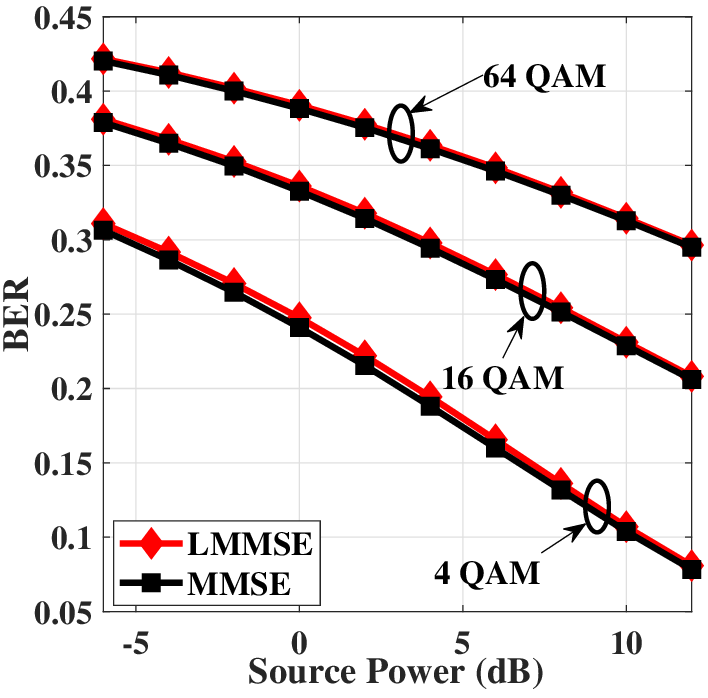}}\\
    \subfloat[]{\label{t11}\includegraphics[width=65mm,height=55mm]{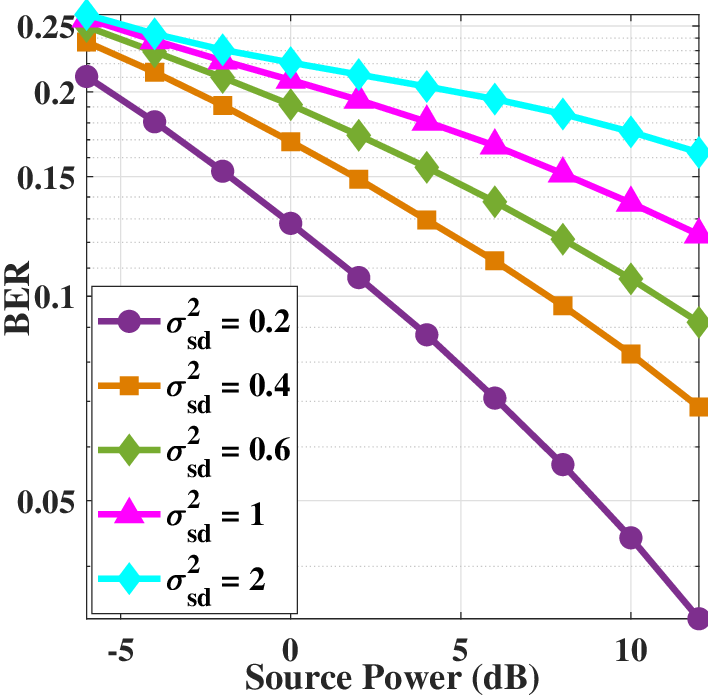}}  \hspace{10mm}
	\subfloat[]{\label{t22}\includegraphics[width=65mm,height=55mm]{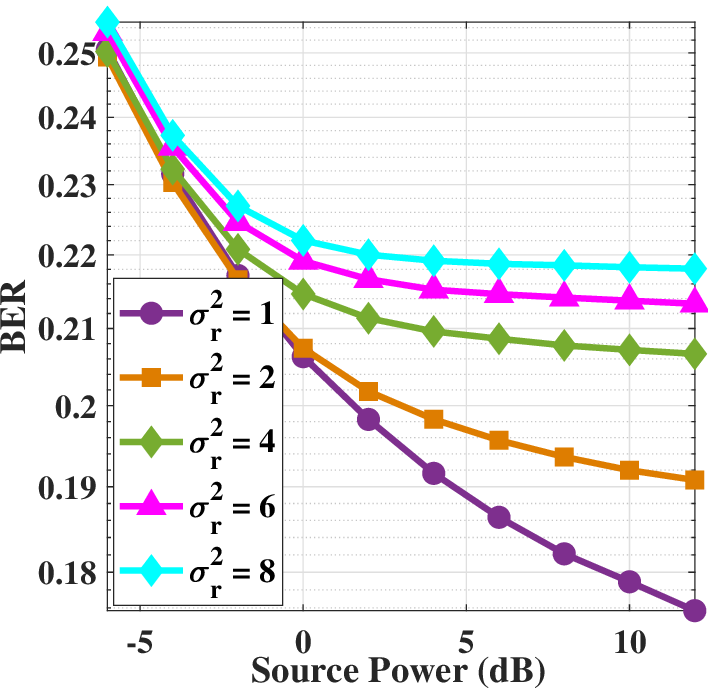}}
	\caption{The BER versus source power is shown for the SISO DCO-OFDM ISDSN-RIN VLC system for (a) all estimators under BPSK modulation; (b) $4, 16,$ and $64$ QAM for LMMSE and MMSE estimators; (c) for different values of \( \sigma^2_{sd}\) for the MMSE estimator; (d) for different values of \( \sigma^2_{r}\) for the MMSE estimator.}
	\label{P1}
    \vspace{-1mm}
\end{figure*}
A similar trend is evident in Fig. \ref{R1w}\subref{sh3}, which demonstrates the impact of RIN noise on the estimation accuracy of various techniques within the DCO-OFDM ISDSN RIN VLC system, while maintaining fixed values for $\sigma^2_{w} = 0.005$, $N=1$, and $\sigma_{sd}^2 = 0.01$. An overall MSE increase is observed across all methods when $\sigma_r^2 = 0.3$ is compared to $\sigma_r^2 = 0.15$, with the MMSE approach consistently achieving the lowest MSE across the evaluated scenarios. This is indeed anticipated, since we directly minimize the MSE. This emphasizes the detrimental effect of high $\sigma_r^2$ on estimation accuracy. Additionally, the discrepancy in performance among the estimators becomes more pronounced for higher $\sigma_r^2$. Nonetheless, at elevated power levels and lower RIN noise, the MSE values of all techniques exhibit convergence. These observations collectively affirm the robustness of the proposed MMSE strategy in achieving reliable estimation across a range of RIN noise conditions.

Fig. \ref{R1w}\subref{sh4} presents the relationship between MSE and the number of transmitted DCO‑OFDM frames $N$ for $\sigma_w^2 = 0.01$, $\sigma^2_{sd} = 0.01$, and $\sigma_r^2 = 0.1$. As $N$ increases, from $N=2$ to $N=4$, all estimators exhibit a marked MSE reduction, with the MMSE approach consistently producing the smallest estimation error. Moreover, the performance differential among the estimators diminishes as more frames are transmitted. These findings collectively validate the superior performance of the proposed MMSE technique under simultaneous ISDSN, RIN, and thermal noise conditions.

Fig. \ref{P1}\subref{p21} illustrates the BER performance of the proposed MMSE-based estimator alongside alternative techniques for the DCO-OFDM ISDSN RIN VLC system, evaluated under fixed noise variances $\sigma_w^2 = 0.01$, $\sigma_{sd}^2 = 1$, $N = 1$, and $\sigma_r^2 = 0.01$ for BPSK modulation. Naturally, the ISDSN and RIN dramatically reduce the performance of the receiver. The considerably lower BER of our MMSE-based CE in comparison to other methods indicates the enhanced estimation accuracy. Moreover, the proposed MMSE-based scheme approaches the BER performance of a receiver relying on perfect CSI (PCSI),
illustrating accurate signal recovery.

Fig. \ref{P1}\subref{p231} portrays the BER performance of the proposed MMSE and LMMSE-based estimators for different QAM modulation orders ($4$, $16$, and $64$) in a DCO-OFDM ISDSN-RIN VLC system, evaluated under fixed noise variances of $\sigma_w^2 = 0.01$, $\sigma_{sd}^2 = 1$, $N = 1$, and $\sigma_r^2 = 0.01$. For all modulation orders, the MMSE estimator consistently achieves a lower BER than the LMMSE estimator, while the BER performance of both estimators degrades as the modulation order increases. These observations collectively confirm the robustness and efficiency of the proposed MMSE scheme in achieving reliable performance across different modulation orders.


In Fig. \ref{P1}\subref{t11}, the BER performance is analyzed for different ISDSN variances using the proposed MMSE estimator, while keeping the thermal noise and RIN variances fixed at $\sigma_w^2 = 0.01$ and $\sigma_r^2 = 0.01$. The results show that ISDSN improves significant BER degradation on the DCO-OFDM VLC system, with the impact becoming increasingly severe as the ISDSN variance escalates. A similar behavior is observed in Fig. \ref{P1}\subref{t22}, where the RIN variance is increased from $1$ to $8$ under fixed noise conditions $\sigma_w^2 = 0.01$ and $\sigma_{sd}^2 = 0.01$ for the MMSE estimator. In this scenario, RIN also causes notable BER deterioration, and the magnitude of degradation intensifies with higher RIN variance. These observations collectively demonstrate that as expected, both ISDSN and RIN adversely affect the estimation performance of the considered system, while highlighting the robustness of the proposed MMSE estimator in maintaining reliable estimation accuracy.
\section{Conclusions}
The joint effects of RIN and ISDSN were investigated, together with thermal noise, on an LD‑based indoor O-OFDM VLC system subject to a random channel. RIN was modeled as being directly proportional to the transmitted optical power and channel gain, whereas ISDSN was assumed to scale with the square root of both. We derived the Fisher information matrix for the system model to compute the BCRLB for the stochastic VLC channel contaminated by RIN and ISDSN contributions. Closed‑form MSE expressions are then obtained for the LS, ML, MAP, MMSE, and LMMSE estimators, under combined RIN and ISDSN influence and validated via extensive simulations. The results revealed pronounced performance degradation induced by ISDSN and RIN, and confirmed that, as expected, the MMSE estimator consistently yields the lowest MSE. A promising future direction is to formulate the joint PDF of the combined LoS and position-dependent microscopic NLoS components and to assess the resulting CE performance under the combined impact of RIN and ISDSN.
\bibliographystyle{IEEEtran} 
\bibliography{citation1}

\end{document}